\documentclass[]{interact}
\usepackage{epstopdf}

\usepackage{natbib}
\bibpunct[, ]{(}{)}{;}{a}{}{,}
\renewcommand\bibfont{\fontsize{10}{12}\selectfont}

\theoremstyle{plain}

\theoremstyle{definition}

\theoremstyle{remark}

\usepackage{multirow}
\usepackage{amsmath}
\usepackage{amsfonts}
\usepackage{amssymb}
\usepackage{graphicx}
\usepackage{textcomp, gensymb}
\usepackage{caption}
\usepackage{subcaption}
\usepackage{placeins}
\usepackage{float}
\usepackage{xcolor}
\usepackage{gensymb}
\usepackage{adjustbox}
\usepackage{booktabs}
\usepackage{pdflscape}
\usepackage{longtable}

\newcommand\numberthis{\addtocounter{equation}{1}\tag{\theequation}}

\usepackage[colorlinks=true, 
            linkcolor=black,
            citecolor=black,
            filecolor=black,
            urlcolor=black]{hyperref}

\begin{document}

\title{Understanding the Impact of Seasonal Climate Change on Canada's Economy by Region and by Sector}

\author{
\name{Shiyu He\textsuperscript{a,$*$}\thanks{*Email: s35he@uwaterloo.ca}, Trang Bui\textsuperscript{a}, Yuying Huang\textsuperscript{a}, Wenling Zhang\textsuperscript{a}, Jie Jian\textsuperscript{a}, Samuel W.K. Wong\textsuperscript{a}, and Tony S. Wirjanto\textsuperscript{a}}
\affil{\textsuperscript{a}Department of Statistics and Actuarial Science, University of Waterloo}
}



\maketitle

\begin{abstract}
    To assess the impact of climate change on the Canadian economy, we investigate the relationship between seasonal climate variables and economic growth across provinces and economic sectors. We also provide projections of climate change impacts up to the year of 2050, taking into account the diverse climate change patterns and economic conditions across Canada. Our results indicate that rising Winter temperature anomalies have a notable adverse impact on Canadian economic growth. Province-wide, Quebec, Manitoba, and Ontario are anticipated to experience larger negative impacts, whereas British Columbia is less vulnerable. Industry-wide, Finance and Real Estate, Science and Technology, and Information, Culture and Recreation are consistently projected to see mild benefits, while adverse effects are predicted for Manufacturing, Agriculture, and Mining. The disparities of climate change effects between provinces and industries highlight the need for governments to tailor their policies accordingly, and offer targeted assistance to regions and industries that are particularly vulnerable in the face of climate change. Targeted approaches to climate change mitigation are likely to be more effective than one-size-fits-all policies for the whole economy.
\end{abstract}


\section{Introduction} 

In recent decades, climate change has had far-reaching impacts on a global scale, including in Canada \citep{arnell2019global, greaves2021climate, houghton1989global, karl2003modern, kovats2005climate, lemmen2007impacts}. Recent shifts in temperature and precipitation patterns may have increased the frequency and magnitude of severe environmental events like droughts, floods, and wildfires, causing tragedies for countless families and communities \citep{abatzoglou2016impact,hirabayashi2008global,luber2009climate,rodrigues2023drivers}. Meanwhile, climate change has also been found to have impacts on the economy \citep{burke2015global, stan2021financial}. For example, the average cost of weather-related disasters and catastrophic losses each year has risen to the equivalence of over 5\% of annual economic growth in Canada over the last decade \citep{canadianclimateinstitute2024drought}. Further, rising temperature is found to be associated with decreasing productivity \citep{burke2015global}; and the intensifying climatic risks escalate the cost of capital for developing countries, exacerbating their governmental debt burdens \citep{UN2018climate}. 
With the world witnessing a multitude of human activities that are likely accelerating climate change \citep{trenberth2018climate,verheyen2005climate}, it is not only necessary but also urgent that the impacts of climate change are thoroughly studied. In this study, we focus on Canada as a case study to shed light on the broader implications of climate change on economies. 

Canada's heterogeneous geographical and economic patterns make it an instructive case study to understand the diverse challenges that climate change poses.
Being the second-largest country by land area, Canada has a diversity of climate zones ranging from the mild and humid maritime weather in British Columbia to the cold arctic weather in the Northwest Territories \citep{oke1998surface}. During the period 1948-2012, Canada witnesses a prevalent warming trend in winters, with western regions (northern British Columbia and Alberta, Yukon, Northwest Territories, and western Nunavut) exhibiting the most pronounced effects. Meanwhile, larger variations are found in precipitation patterns: whereas Northern Canada (Yukon, Northwest Territories, Nunavut, and northern Quebec) experienced increased precipitation across all seasons, decreasing trends in winter precipitation are dominant in the Southwest of Canada (British Columbia, Alberta, and Saskatchewan) \citep{vincent2015Observe}. 


Besides the varying climatic patterns, Canadian economic landscapes also differ by region and by sector due to the diversity of natural conditions and cultural preferences. Alberta's economy, for instance, is anchored in its mining industry. On the contrary, Ontario's economy relies heavily on finance and real estate \citep{statcanGDP}. Hence, the economic implications of climate change in Canada widely vary across provinces and industries.

There is an extensive literature on climate change and its economic impacts on Canada. In agriculture, \cite{weber2003regional} offer an overview of the climate change impacts on Canadian agriculture. Regionally, \cite{kulshreshtha2011climate} address the uncertainties brought by warmer climates to agriculture in the Prairie Provinces of Canada, while \cite{reza2022impact} study the impact of climate change on crop production and food security in Newfoundland and Labrador. In the case of mining, \cite{prowse2009implications} review the impact of the changing climate on the mining industry in Northern Canada, and \cite{pearce2011climate} characterize the vulnerability of the Canadian mining industry to climate change. Other industries that have been studied include recreation and tourism \citep{dodds2009canada, hewer2018thirty, scott2007climate, scott2019differential}, finance and real estate \citep{greaves2021climate, moudrak2019ahead, williams2013climate}, manufacturing \citep{kabore2023manufacturing}, and utilities \citep{cool2019climate}.
However, these studies often focus on a specific region or economic sector of Canada, and are not specifically designed to provide a holistic picture of climate change in a broader economic context. A comprehensive assessment of the climate change impacts on all economic industries across every province of Canada can be beneficial in that it facilitates comparison between different sectors and provides a bigger picture. This can inform proactive and targeted strategies to address climate change in the Canadian economy.



As to the economic impact of climate change in Canada, there seems to be no clear consensus expressed in the existing studies. 
In terms of manufacturing, various studies have identified a negative impact of climate change, although they examine these impacts from different perspectives. For instance, \cite{OCCI2015climate} express concerns about the disruptions in supply chains due to a climatic effect-induced extreme weather in Ontario. In addition, \cite{kabore2023manufacturing} find that the increasing temperature anomalies in Canada have a detrimental effect on manufacturing output through decreased labour productivity and product demand. These distinct perspectives collectively show the multifaceted challenges posed by climate change to the manufacturing sector in Canada.
Meanwhile, conflicting views are also reported on the climatic impacts on mining. \cite{pearce2011climate} underscore the adverse impacts of climate-related events on mines in Canada and point to climate change as a growing threat to the Canadian mining sector. On the contrary, \cite{prowse2009implications} and \cite{mac2021climatechange} explain the advantages of rising temperature for the mining industry, which includes an increasing number of accessible mines and various other favourable factors.
As for agriculture, \cite{weber2003regional} and \cite{ochuodho2016economic} find possible benefits in most provinces. This stands in contrast to other climate-economic studies that report mixed, and often more pessimistic, findings for the agriculture sector \citep{williams1988estimating,dolan2008climate, kulshreshtha2011climate, pittman2011vulnerability}. Besides manufacturing, mining, and agriculture, climate-economic studies focusing on other industries also draw mixed conclusions \citep{adekanmbi2023assessing, elsasser2002climate, gilmour2010ontario, kulshreshtha2011climate, qian2019climate, scott2007climate}. 
This calls for further studies to draw firmer conclusions about climate change impacts on each economic sector in Canada. 



Our article takes a comprehensive approach to studying the relationship between climate change and the Canadian economy. In particular, we provide estimates and comparisons of climate impacts on each economic sector within each Canadian province. These estimates offer a holistic yet granular perspective on how climate change impacts vary by sector and by region across Canada. Furthermore, while many existing studies primarily focused on annual temperature as the key climate change indicator \citep{burke2015global, chen2019temperature, dell2009temperature}, our modeling approach incorporates seasonal climatic trends in both temperature and precipitation. This consideration helps us better understand the intricate relationship between climate change and economic outputs in the presence of seasonality. 
Finally, based on the results obtained in our study, we make predictions about the climate change impacts on the Canadian economy under various climate scenarios. Separate predictions are made for different economic sectors and across provinces. These predictions enable us to understand the varying susceptibilities of different provincial and economic sectors to climate change, leading us to more comprehensive implications in sustainable finance, future planning, and policies in Canada. 

The rest of the article is organized as follows. In Section 2, we describe in detail how we collect and pre-process our data. A thorough review of our methodology is discussed in Section 3. In Section 4, we present our main findings and results, which include (i) an overview of Canadian climate and economy, (ii) current climate-economic patterns implied by our results, and (iii) projections of climate change impacts for each economic sector in each province in Canada. Finally, we summarize our conclusions, policy implications, and avenues for future studies in Section 5. 

\section{Data}




To disentangle the intricate relationship between climate change and the Canadian economy, in this paper, we rely on two primary types of data: weather data and economic data. The data are collected over the period 1998-2017 from various public websites. The economic data include Canadian general and industry-specific gross domestic product (GDP), annual target interest rate, annual provincial unemployment rates; World Bank's world GDP and commodity price indices for energy and non-energy sectors; and indicators for major economic events. To conduct our analysis, we pre-process our data by (i) transforming the GDP data into GDP per capita growth rate (PCGR), (ii) taking the log differences of the economic indices, and (iii) calculating seasonal climate deviations from the average. The following sections discuss the details of such procedures. 


\subsection{Economic Data}

\subsubsection{GDP}

We extract the raw annual GDP of fifteen economic sectors in ten provinces\footnote{The sparsely populated northern territories of Yukon, Northwest Territories, and Nunavut are excluded from this study.} over 20 years from 1997 to 2017 from \cite{statcanGDP}. To adjust the dollar amounts for inflation over time, we download the GDP data at base prices chained to 2012 dollars. Annual population data in each province during the same period is also collected \citep{statcanPopulation}. For each industry, the corresponding GDP per capita $Y_{p,t}$ of province $p$ in year $t$ is calculated by dividing the GDP by the population.
 
We follow the conventional practice in economic literature to analyze economic patterns by using the GDP per capita growth rates (GDP PCGR) instead of raw GDP data \citep{burke2015global,newell2021gdp}. In particular, the GDP PCGR of province $p$ in year $t$ is calculated as the difference in natural logarithms of annual GDP per capita between year $t$ and year $t-1$,
\begin{equation}
    \text{PCGR}_{p,t} = \log\left(\frac{Y_{p,t}}{Y_{p,t-1}}\right) \approx \frac{Y_{p,t}}{Y_{p,t-1}} - 1.
    \label{eq:PCGR_approx}
\end{equation}
The use of GDP PCGR is justified as it is scale-invariant, thus enabling effective comparisons across different provinces, industry sectors, and time periods. In addition, it is also weakly stationary, which ensures that our regression models involve only weakly stationary variables.

\subsubsection{Other Economic Indices} \label{sec:econ-indices}

In addition, we include a range of economic indices alongside the standard control variables to aid in modelling the regression relationship between climate and economic outcomes. These indices capture fluctuations in commodity markets, labor market conditions, monetary policy, global economic activity, and discrete macroeconomic shocks. To ensure stationarity and consistency in the analysis, all series are transformed to annual growth rates via log differences unless otherwise specified.


Global economic conditions play a critical role in shaping Canada's macroeconomic environment through trade and capital flows. As a proxy for general global economic conditions, we collect world GDP data from World Bank \citep{worldbank_gdp_current}, which is defined as the total monetary value of all goods and services produced worldwide within a year \citep{barro1997macroeconomics}. In addition, commodity markets are also important drivers of the Canadian economy, especially in resource-intensive provinces. To capture these effects, we use annual nominal commodity price indices from the World Bank's Global Commodity Markets dataset, commonly referred to as the Pink Sheet \citep{worldbank_pinksheet}. We include two broad aggregates, specifically, energy and non-energy commodities, which reflect Canada's main trade and resource sectors. These indices are particularly relevant for resource-intensive economies such as Canada, and are widely employed in climate-economy research to capture external terms-of-trade shocks \citep{cashin2014differential,burke2015global}.

To accommodate national factors and the possible role of monetary policy, we include in our study the annual average of the Canadian target interest rate, as provided by the Bank of Canada \citep{boc_rates}. The interest rate is a central policy lever used to influence aggregate demand, with lower rates generally associated with increased consumption and investment \citep{imf1983}.

In terms of local factors, we utilize unemployment rate data as a proxy for labor market dynamics and an indicator of provincial economic performance and socioeconomic vulnerability, both of which are relevant in assessments of climate sensitivity \citep{deschenes2007economic}. We collect annual provincial unemployment rates for individuals aged 25 to 54, regardless of educational attainment or gender, as published by Statistics Canada \citep{statcan_unemployment}. 

Finally, we construct indicator variables for major exogenous economic events that may impact Canadian provinces during the study period from 1997 to 2017. Events are identified through a systematic review of federal government archives, official reports from the Bank of Canada and Statistics Canada, and prominent Canadian news sources. We select events on the basis of their economic significance and predominantly exogenous, province-specific shock profiles. A comprehensive listing of these major events, along with the provinces affected, is presented in Appendix~\ref{app:econ-events}.

\subsection{Climate Data}

From the \cite{climate}, we obtain monthly mean temperature (in Celsius degrees) and total precipitation (in millimeters) data from weather stations across each Canadian province. As temperature and precipitation patterns vary by season, we opt to employ seasonal weather data in our climate-economic model. Such seasonal data (i.e., Spring, Summer, Fall, Winter\footnote{Seasons are defined as Spring (March, April, May), Summer (June, July, August), Fall (September, October, November) and Winter (December, January, February).}) are calculated by taking the average of monthly weather measurements for each season. Note that some weather stations have high missingness in their records, the use of which may affect the robustness of our data aggregation results. To resolve this issue, we remove stations with incomplete measurements and ensure that for each province, within each season, we have sufficient weather stations with complete data to calculate the averages. Further details on the weather data processing and the post-processed spatial distribution of the retained stations are provided in Appendix \ref{app:weather-station}.

In this article, instead of absolute weather measurements, we use weather anomalies, which are defined as the deviations of weather measurements from the corresponding long-term averages over the whole observation period (baseline). In the literature, weather anomalies are often employed to analyze climate change. For example, \cite{rahmstorf2015exceptional} and \cite{lehmann2015increased} used temperature anomalies and precipitation anomalies, respectively, to explain changing phenomena associated with global warming. \cite{kahn2021} investigated how productivity responds to deviations in temperature and precipitation from their long-term moving average historical norms and found that per-capita real output growth is negatively affected only by temperature anomalies. According to \cite{NCEI}, weather anomalies are preferable to absolute weather measurements because they allow more meaningful comparisons between locations and better illustrations of weather trends. When computing the average absolute weather variables, many factors, such as the variations in locations, elevation, and data collection styles of weather stations, can largely affect the results. 
In contrast, the baseline reference for anomaly is less affected by single factors, and the anomaly calculation will remain relatively robust. In this study, we adopt the period of 1998-2017 as the baseline reference period rather than the conventional climatological normal period (1981-2010) due to data availability constraints. We acknowledge that this choice of baseline may introduce minor look-ahead biases and sensitivities in terms of projection and results interpretation in our study. Nevertheless, our computations of the temperature and precipitation anomalies are formulated as follows: 

\begin{equation}
    \text{TempAnomaly}_{p,s,t}=\text{Temp}_{p,s,t}-\overline{\text{Temp}}_{p,s}=\text{Temp}_{p,s,t}-\dfrac{1}{20}\sum_{t=1998}^{2017}\text{Temp}_{p,s,t},
    \label{eq:temp_anomaly}
\end{equation}

\begin{equation}           
    \text{PrecAnomaly}_{p,s,t}=\dfrac{\text{Prec}_{p,s,t}-\overline{\text{Prec}}_{p,s}} {\overline{\text{Prec}}_{p,s}} \times 100 =\dfrac{\text{Prec}_{p,s,t}-\dfrac{1}{20}\sum_{t=1998}^{2017}\text{Prec}_{p,s,t}}{\dfrac{1}{20}\sum_{t=1998}^{2017}\text{Prec}_{p,s,t}}\times 100,
    \label{eq:prec_anomaly}
\end{equation}
where $\text{TempAnomaly}_{p,s,t}$ (in Celsius degrees) is the deviation of the seasonal average temperature in province $p$, season $s$, year $t$ from its sample average. Similarly, $\text{PrecAnomaly}_{p,s,t}$ (in percentage) is the relative percentage change in seasonal average precipitation in province $p$, season $s$, and year $t$ compared to its sample average. 

\section{Methods}

\subsection{Model Fitting and Selection} \label{sec:model}


To assess the influence of climate variables on GDP PCGR, we employ a series of linear mixed-effects models  \citep{west2022linear}. These models extend the standard linear regression by allowing the effects of independent variables to be treated as either fixed or random. In particular, the effect of a variable is modeled as fixed if the observed values in the sample data exhaust the possible values in the population, and the effect of each of these values is interesting in itself. On the other hand, random effects are used to model a variable whose observed values are considered to be only a part of the population, and the goal is to understand such an underlying population \citep{gelman2005analysis, searle2009variance}. In our case, we model the effect of Province as fixed because our data set encompasses ten major provinces in Canada. The objective of our study is to understand the GDP PCGR patterns in each of these provinces. The effect of Year, strictly speaking, can be considered random, since our data do not cover all past and future years, and the impact of a given year may vary in unpredictable ways due to unobserved social, political, or economic events. It is worthwhile to stress that our goal in this study is not to estimate the effect of any particular year. However, in the economics literature, it is standard to model Year as fixed in two-way fixed effects models, since this approach does not require the other regressors to be independent of the year effect \citep{wooldridge2010econometric,hsiao2022analysis}. To examine this subtlety, we consider three ways to treat the variable Year. The first is to consider the variable Year as a set of fixed effects in two-way fixed effect models. The next is to omit the variable Year entirely, allowing us to focus on other sources of time-related variation, such as other economic factors. Finally, we consider incorporating the variable Year as a set of random effects, which is particularly useful when annual economic indices are included, since some indices remain constant within a year and would otherwise be perfectly collinear with fixed year effects. Overall, the linear mixed-effects models allow us to appropriately account for possible unobserved confounding factors that may influence the GDP PCGR across time and across provinces. In addition, the use of linear models instead of other machine learning models \citep{stan2021financial} ensures the high degree of interpretability of our results.

While province fixed effects accounts for time-invariant heterogeneity between provinces, and GDP per capita controls for broad economic output, our model must also account for dynamic, province-year-specific shocks that could potentially confound the relationship between climate and economic outcomes \citep{kahn2021,tol2018economic}. Such shocks may include major policy changes, external economic events, or major structural breaks. To address this issue, in some specifications, we consider incorporating as fixed effects global and local economic indices as discussed in Section~\ref{sec:econ-indices}, to capture both external and local economic conditions that may influence GDP PCGR. The use of log differences in these indices removes time trends and makes these economic indices weakly stationary. Additionally, we account for major economic events, such as recessions, policy shifts, or financial crises, in some of our regression specifications. These events can influence economic outcomes differently across provinces and over time. To capture this heterogeneity without overfitting the model, we treat event indicators as random effects. Finally, we consider province-specific time trends or province-specific random effects, allowing for heterogeneous long-term structural changes and sensitivities across regions. Overall, while these additions help mitigate potential bias, we acknowledge that other relevant factors, such as technological change, policy reforms, trade dynamics, and sectoral shifts, are not explicitly modeled due to data limitations. Following the standard practice in the climate-econometric literature \citep{dell2014we}, we view our robustness checks of including these economic indices as a reasonable way to showcase the credibility of the estimated climate effects, while recognizing that any remaining omitted variable bias is unaccounted for in our study.

In terms of modeling the effects of climate variables, \cite{burke2015global} point out that temperature has a quadratic relationship with productivity and thus economic activity. In particular, they hypothesize that productivity increases with temperature up to a certain optimal point and then decreases, exhibiting the law of diminishing returns. Such a hypothesis can be rationalized as neither extremely cold nor extremely hot weather is good for human activities. On the other hand, \cite{newell2021gdp} raise concerns that the quadratic model may or may not work for different countries and for different economic states. Hence, for a robustness check, we consider both linear and quadratic effects of the seasonal mean temperature ($\text{Temp}_{p,s,t}$) and absolute seasonal precipitation ($\text{Prec}_{p,s,t}$) anomalies in our model fitting. By decomposing annual temperatures and precipitation into seasonal components, we aim to capture and quantify the effects of varying seasonal climate change patterns on GDP PCGR. Finally, the first lag of the GDP PCGR is also added to our models to control for any remaining dynamic effects. 

Altogether, our models are specified as
\begin{align*}
    \text{PCGR}_{p,t} & = \alpha_0 + \alpha_p + \eta_t + \gamma \text{PCGR}_{p,t-1} + \beta^{\top} \text{Climate effects} + \delta^{\top} \text{Economic indexes} + \epsilon_{p,t}, \numberthis \label{eq:main_model}
\end{align*}
where $\alpha_p$ is the fixed effect for province $p$, $\eta_t$ is the effect for year $t$. In addition, $\beta$ represents the vector of coefficients for climate variables and $\delta$ represents the vector of coefficients for economic indices, if included in a model specification. We fit different specifications of the general model in Equation~\eqref{eq:main_model} to the overall GDP PCGR\footnote{The overall GDP PCGR is calculated by~\eqref{eq:PCGR_approx} where $Y_{p,t}$ represents overall GDP per capita aggregated across all economic sectors. This is distinct from the GDP PCGR for each economic sector.} and evaluate their goodness of fit using both Akaike Information Criterion (AIC) and Bayesian Information Criterion (BIC), two popular criteria for model selection \citep{west2022linear}. Note that the AIC tends to favor larger models. For parsimonious modeling, we will rely on the BIC for model selection when the results of AIC and BIC are conflicting. Finally, based on the model fitting results for the overall GDP PCGR, we choose the optimal working model to fit the GDP PCGR data for each economic sector. 

Since we consider linear, quadratic, and interactions of climate variables in our model fitting, to summarize the effects of specific climate variables on GDP PCGR, we use average marginal effects, which are calculated as
\begin{equation}
    \mathbb{E}_t\left[\frac{\partial \text{PCGR}_{p}}{\partial \text{T}_{s,p}}\bigg|_{\text{T}_{s,p} = \text{T}_{s,p,t}, \text{P}_{s,p} = \text{P}_{s,p,t}}\right] \hspace{5mm} \text{and} \hspace{5mm} \mathbb{E}_t\left[\frac{\partial \text{PCGR}_{p}}{\partial \text{P}_{s,p}}\bigg|_{\text{T}_{s,p} = \text{T}_{s,p,t}, \text{P}_{s,p} = \text{P}_{s,p,t}}\right], \label{eq:marginal}
\end{equation}
where T and P respectively denote temperature and precipitation anomalies, and $\mathbb{E}_t$ denotes an average taken over different years. These average marginal effects can be interpreted as the average relative change in GDP PCGR resulting from a one-unit increase in either the temperature anomaly (T) or the precipitation anomaly (P). As we employ linear models, the derivatives in \eqref{eq:marginal} can be calculated in a straightforward manner, whether the climate variables include linear, quadratic terms, or interactions. 

To address potentially correlated errors within each province, we employ robust clustered standard errors to quantify the remaining uncertainty in our model estimates. Furthermore, as we only have data for 10 provinces, we use the Satterthwaite correction for small sample sizes when calculating our p-values \citep{pustejovsky2018small}. 

\subsection{Projection}

Based on the model fitting, we make projections of the climate change impacts on GDP PCGR for each province and industry up to 2050, the year by which the Government of Canada commits to achieve net-zero emissions \citep{netzero}. Our projections are based on different climate-economic scenarios and the corresponding climate predictions by 2050. 

\subsubsection{Climate Projection} \label{sec:climate-projection}

Projections of future climate patterns in Canada up to 2050 are obtained by referring to the Coupled Model Intercomparison Project Phase 6 (CMIP6) Global Climate Models (GCM) \citep{o2016scenario}. These projections are generated under a range of Representative Concentration Pathways (RCPs), which are greenhouse gas (GHG) concentration trajectories associated with alternative socioeconomic scenarios, as adopted by the Intergovernmental Panel on Climate Change (IPCC) \citep{van2011representative}. Specifically, we use the 50th percentile, i.e., ensemble median, of the projections in RCP2.6, RCP4.5, and RCP8.5 pathways, which correspond, respectively, to low-, intermediate-, and high-emissions scenarios. The inclusion of multiple RCPs allows for a comprehensive assessment of potential climate-economic outcomes for Canada, taking into account short-term natural climate variability, model ensemble uncertainties, and the long-term effects associated with different emissions trajectories.

Specifically, we download the projected climate data for the different emission scenarios from Canadian Climate Data and Scenarios \citep{canada2019cmip6}. The data consists of projected changes in mean temperature ($^\degree$C) and total precipitation (\%) for the near-term (2021-2040) and mid-term (2040-2060) compared to the reference period 1995-2014. We use these projections to extrapolate the future seasonal temperatures and precipitations ($\text{Temp}_{s,p,t}^{\text{pred}}$ and $\text{Prec}_{s,p,t}^{\text{pred}}$) by assuming linear climatic growths during the periods 2014-2040 and 2040-2060. The corresponding future climate anomalies are then calculated according to \eqref{eq:temp_anomaly} and \eqref{eq:prec_anomaly}.

\subsubsection{Projection of Climate Change Impact on GDP}

To quantify the effect of climate change on GDP in the future, we compare the projected GDP per capita under with- and without-climate-change scenarios. In the without-climate-change scenario, GDP per capita is assumed to grow at the same rate as the average historical growth over the data period ($\overline{\text{PCGR}}_{p,1998-2017}$). In the with-climate-change scenario, we instead use the climate projections discussed in Section \ref{sec:climate-projection} as inputs to the working model in Section \ref{sec:model} to obtain the predicted values of PCGR in each future year ($\text{PCGR}_{p,t}$). 

According to the PCGR calculation in Equation \eqref{eq:PCGR_approx}, the GDP per capita of province $p$ at a year $t$ is
\[Y_{p,t} = Y_{p,t-1} \times e^{\text{PCGR}_{p,t}}.\] 
Hence, the projected climate change impact in province $p$ by year $T$, i.e., the difference in GDP per capita between the with- and without-climate-change scenarios up to year $T$ can be calculated as
\begin{align*}
    \% \Delta Y_{p,T} & = \frac{Y_{p,T,\text{with CC}}}{Y_{p,T,\text{without CC}}} - 1 \\
    & = \exp\left(\sum_{t=2018}^{T} \left[ \widehat{\text{PCGR}}_{p,t} - \overline{\text{PCGR}}_{p, 1998-2017}  \right]\right) - 1, \numberthis \label{eq:impact}
\end{align*}
where $\widehat{\text{PCGR}}_{p,t}$ denote the predicted values and $\overline{\text{PCGR}}_{p, 1998-2017}$ denote the historical baseline. Since we use a linear model for PCGR, the difference $\widehat{\text{PCGR}}_{p,t} - \overline{\text{PCGR}}_{p, 1998-2017}$ in \eqref{eq:impact} reduces to the difference between the climate effect terms in the selected model, when climate anomaly projections vs. average of historical climate anomalies in 1998-2017 are used. We thus can use \eqref{eq:impact} to project the average impacts of climate change on each industry in every Canadian province by 2050, without specific projections for future changes in economic indices or year-specific effects. 


\subsubsection{Projection Uncertainty in Climate Change Impact}

Following \cite{burke2015global}, we use a block bootstrap technique to quantify uncertainty in the climate change impact on GDP. In each replication, 10 provinces are sampled with replacement, and all observations within a selected province are retained to preserve the within-province correlation structure. In this way, the bootstrap draws retain the correlation structure within each province. This procedure is repeated 1,000 times, with the model refitted in each replication to obtain estimates for model coefficients. These coefficients are used to produce sensitivity projection paths that describe model and projection uncertainties. The results for different RCP paths are also presented to account for uncertainty in climate projections.

\section{Results}

\subsection{Overview of the Canadian Economy and Climate Change Patterns}

To assess the economic impact of climate change in Canada, it is important to establish an understanding of the country's broader economic landscape and its evolving climate patterns.

\subsubsection{Industry Shares}\label{sector: econ}

\begin{figure}[H]
         \centering
\includegraphics[width=\textwidth]{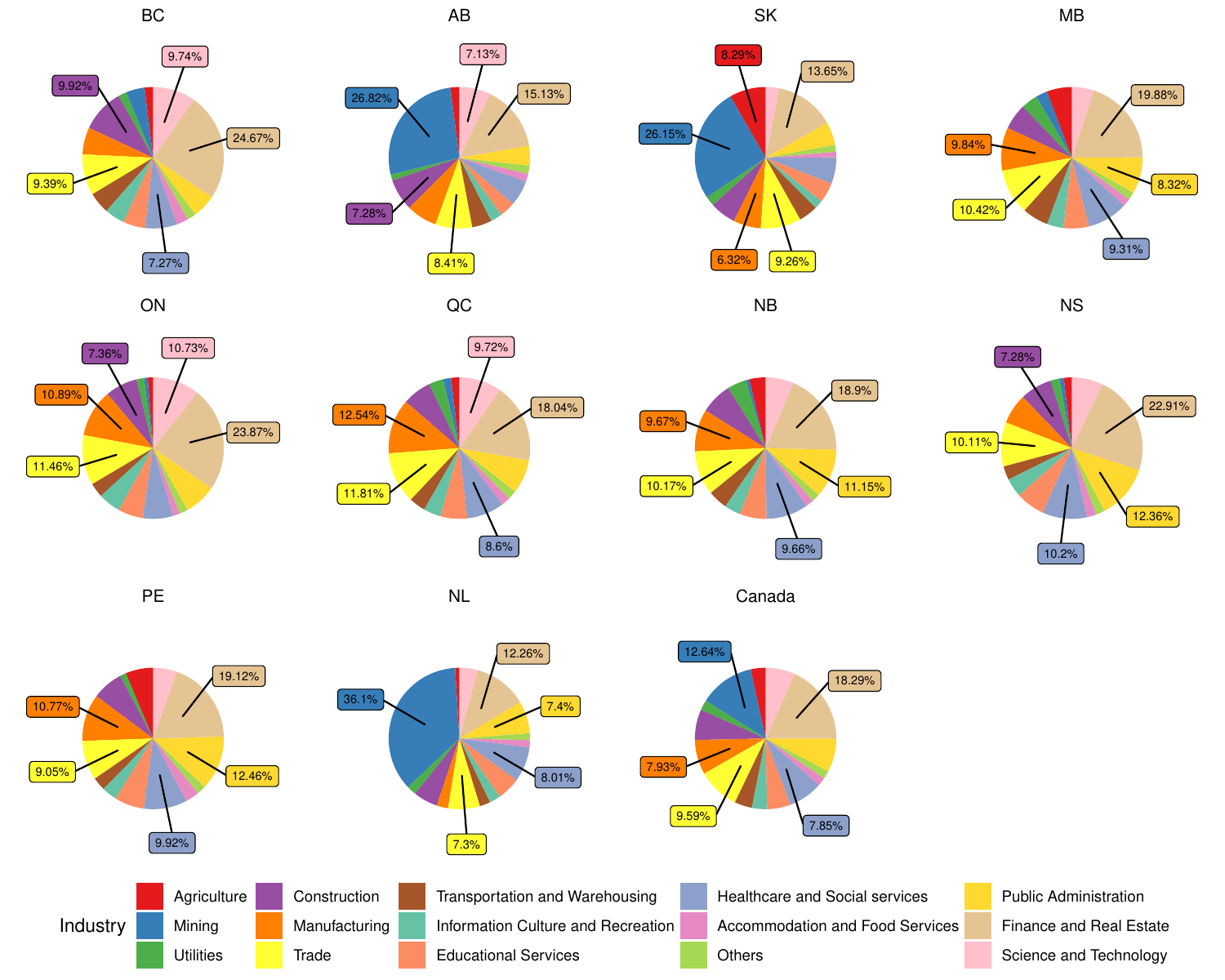} 
         \caption{Top 5 Industries in 2022, Provinces are labelled British Columbia (BC), Alberta (AB), Saskatchewan (SK), Manitoba (MB), Ontario (ON), Quebec (QC), New Brunswick (NB), Nova Scotia (NS), Newfoundland (NL), and Prince Edward Island (PE).}
         \label{overview_gdp_shares}
\end{figure}

We analyze the industrial composition of fifteen industries in 2022 for each province and across Canada, as depicted in Figure~\ref{overview_gdp_shares}. Nationally, the top five industries contributing to the Canadian economy are Finance and Real Estate, Mining, Trade, Manufacturing, and Healthcare and Social Services. In all provinces, Finance and Real Estate, along with Trade, consistently appear among the top five contributing industries. 

As Canada is a country heavily reliant on natural resources, the Mining sector represents one of its most significant industries. In 2021, the Mining industry contributed a substantial 125 billion to the national GDP, constituting 22\% of Canada's total domestic exports and employing 665,000 individuals directly and indirectly \citep{miningCAN}. Another important sector is Manufacturing, as it currently contributes around 174 billion, equivalent to about 10 percent of the total GDP. Canadian manufacturers play a crucial role on the global stage, exporting over 354 billion annually, equivalent to 68 percent of Canada's merchandise exports \citep{manufCAN}. 

Provinces in Canada exhibit diversity in their specialized industries, reflecting their regional differences and uniquely relative economic strengths. For instance, Alberta, Saskatchewan, and Newfoundland and Labrador have a strong presence in the Mining sector, accounting for more than 26\% of the provincial GDP. By contrast, Ontario and Quebec are known for their Manufacturing sectors, with particular strength in automotive and aerospace manufacturing. Agriculture is important in the provinces of Saskatchewan, Manitoba, and Prince Edward Island. Meanwhile, Science and Technology ranks among the top five industries in the four provinces: British Columbia, Alberta, Ontario, and Quebec.

\subsubsection{Distribution of Projected Weather Anomalies}\label{sec: proj_weather}

\begin{figure}[!h]
\captionsetup[subfigure]{labelformat=empty}
    \begin{subfigure}[b]{0.49\linewidth}
         \centering
         \includegraphics[width=\textwidth]{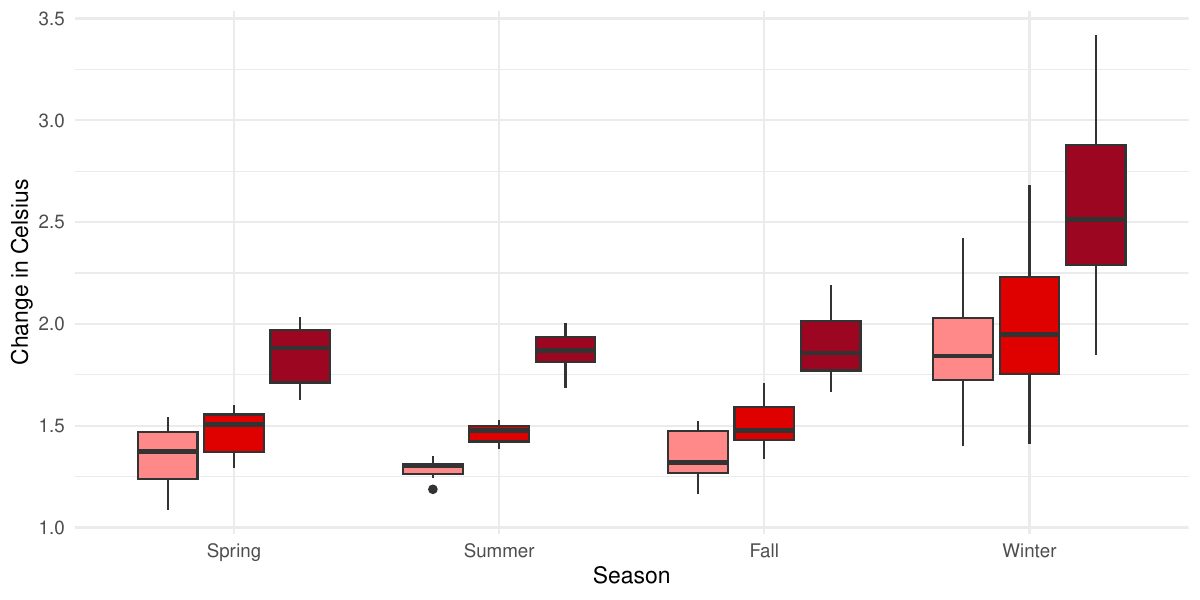} 
         \caption{(a) Temperature anomalies sorted by seasons}
         \label{fig:2050projected_anomalies_1}
    \end{subfigure} 
    \begin{subfigure}[b]{0.49\linewidth}
         \centering
         \includegraphics[width=\textwidth]{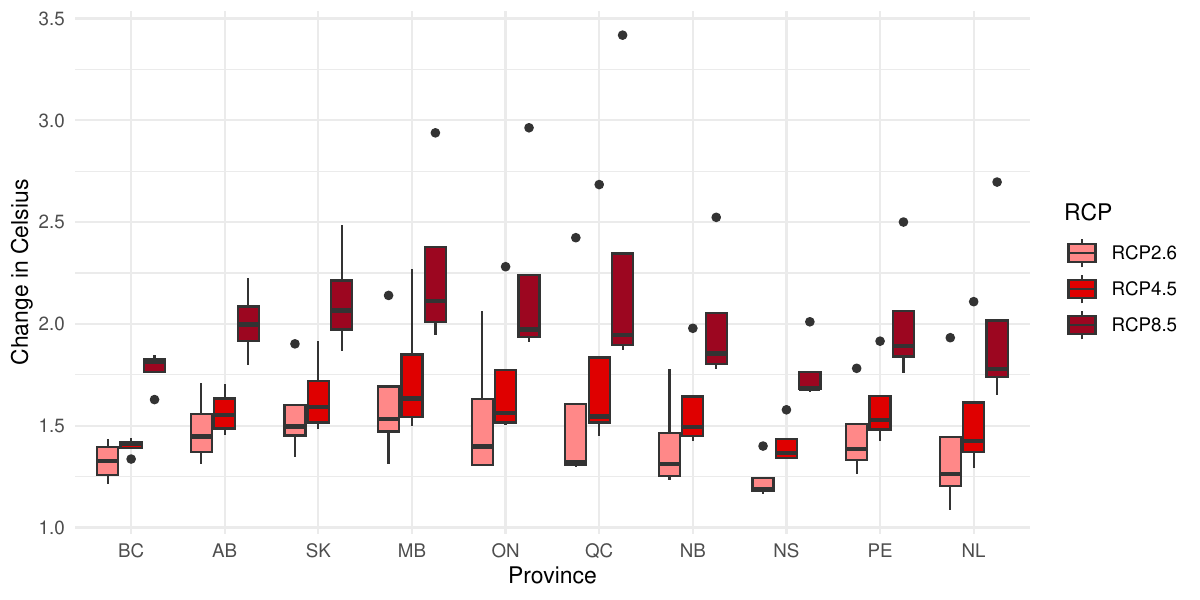}
         \caption{(b) Temperature anomalies sorted by provinces}
         \label{fig:2050projected_anomalies_2}
    \end{subfigure} 
    \begin{subfigure}[b]{0.49\linewidth}
         \centering
         \includegraphics[width=\textwidth]{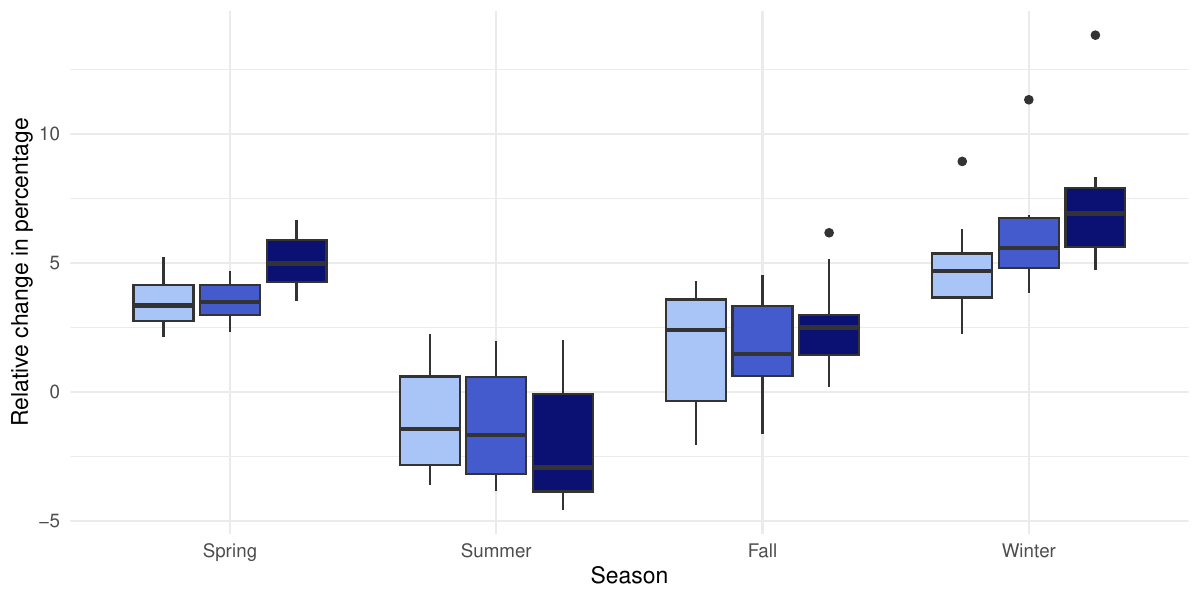} 
         \caption{(c) Precipitation anomalies sorted by seasons}
         \label{fig:2050projected_anomalies_3}
    \end{subfigure} 
    \begin{subfigure}[b]{0.49\linewidth}
         \centering
         \includegraphics[width=\textwidth]{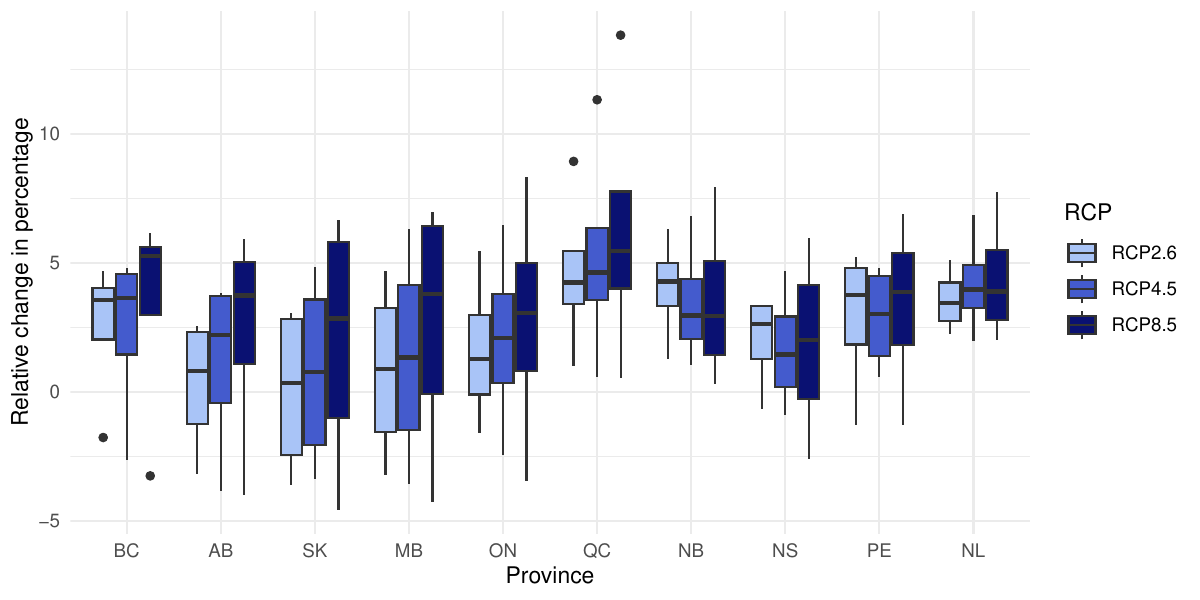}
         \caption{(d) Precipitation anomalies sorted by provinces}
         \label{fig:2050projected_anomalies_4}
    \end{subfigure} 
    \caption{Predicted temperature and precipitation anomalies by 2050 with baseline calculated over the 1998-2017 period.} \label{2050projected_anomalies}
\end{figure}

This section provides a review of projected weather anomalies by 2050 based on three GHG emission scenarios (RCP2.6, RCP4.5, and RCP8.5). Four boxplots are shown in Figure \ref{2050projected_anomalies} to illustrate the distribution of predicted weather anomalies by season and by province, respectively. 

In general, temperature anomaly will increase by at least 1.1 degrees Celsius in 2050 across all seasons and RCP scenarios. However, according to Figure \ref{2050projected_anomalies}\subref{fig:2050projected_anomalies_1}, different seasons will have varying magnitudes and ranges of temperature increments. Among all seasons, Winter is projected to be the most severely affected by global warming, as it is forecasted to have the greatest temperature increase (1.9$^\degree$C under RCP2.6, 2.0$^\degree$C under RCP4.5, and 2.6$^\degree$C under RCP8.5). Figure \ref{2050projected_anomalies}\subref{fig:2050projected_anomalies_1}  also shows that anomalies are predicted to vary across provinces, with greatest variations in Winter and smallest in Summer. 

Figure \ref{2050projected_anomalies}\subref{fig:2050projected_anomalies_2}  provides a breakdown of temperature anomalies across provinces. Notably, Canada's Atlantic region (i.e., New Brunswick, Nova Scotia, Prince Edward Island, and Newfoundland) and British Columbia province will experience smaller temperature increases compared to other provinces. This might be attributed to the ocean's ability to absorb and store heat, thereby tempering the effects of global warming. On the other hand, the Prairie provinces of Manitoba, Saskatchewan, and Alberta are projected to experience greater temperature increases under all RCP scenarios.

The change in seasonal precipitation levels, as illustrated in Figure \ref{2050projected_anomalies}\subref{fig:2050projected_anomalies_3}, also suggests a high degree of variation. While the precipitation levels are projected to decrease in Summer (averaged decrease of 11\% under RCP2.6, 13\% under RCP4.5, and 20\% under RCP8.5), they are expected to increase in all other seasons. Specifically, Spring and Winter are predicted to see greater increases than Fall.

Finally, Figure \ref{2050projected_anomalies}\subref{fig:2050projected_anomalies_4} provides a picture of how precipitation will be affected in the year 2050 for each province. It is generally predicted that precipitation patterns in all prairie provinces (i.e., Alberta, Saskatchewan, and Manitoba) will not drastically change under optimistic scenarios (under RCP2.6 or RCP4.5). In contrast, all other provinces are expected to receive more precipitation in the future.

Overall, the analysis of predicted temperature and precipitation shows varying patterns across seasons and provinces. This supports our choice of season-specific climate-economic modeling. 

\FloatBarrier

\subsection{Current Patterns of Climate Change Impact on Canadian Economy}


As discussed in Section \ref{sec:model}, in order to understand the current relationship between climate and GDP growth in Canada, we fit different specifications of the linear-mixed effects model in Equation~\eqref{eq:main_model}, which are described as follows. Model (1) is the traditional two-way fixed effects model. Model (2) expands upon this by including both linear and quadratic terms of climate anomalies. As the quadratic terms in Model (2) do not improve the BIC results, we use linear terms of climate anomalies in the remaining models. Next, Model (3) considers adding economic indices while removing the year effect, as these variables would be perfectly collinear in a fixed effect model. Model (4) adds to Model (3) the indicator variables for significant economic events. Finally, Models (5) and (6) repeat Models (3) and (4), respectively, with Year added back to the models as random effects. The results of these model specifications are summarized in Table \ref{tab:model_selection}. We also conduct sensitivity analyses in which we consider interactions of climate variables and province-specific trends; the results of those model fits are included in Appendix \ref{app:sensitive-model-spec}. 

Table~\ref{tab:model_selection} shows that Winter temperature is significantly associated with GDP PCGR in Models (1), (2), (5), and (6). Across all specifications, the estimated effects of Winter temperature anomalies are negative, implying that rising Winter temperatures could be associated with a decrease in Canadian GDP growth. In contrast, other climate variables, namely Spring, Summer, and Fall Temperature, together with Precipitation in general, do not exhibit either notable consistent patterns or significant associations based on our model-fitting. 
Besides the results shown in Table~\ref{tab:model_selection}, we also observe similar patterns while fitting other model specifications or using different climate data calculations (See Appendices~\ref{app:sensitive-model-spec} and \ref{app:sensitive-model-climate}, respectively). This consistency strengthens the robustness of our results. Overall, as Model (5) produces the smallest value of BIC and a low AIC, we choose Model (5) as our working model and use it to obtain results for GDP PCGR in different industrial sectors. 


\begin{table}[htbp]
\scriptsize
\centering
\begin{tabular}{lcccccc}
  \hline
   & (1) & (2) & (3) & (4) & (5) & (6) \\ 
  \hline
  \textbf{Model specification} & & & & & & \\
  Year Effect & Fixed & Fixed & No & No & Random & Random \\ 
  Economic Indices & No & No & Yes & Yes & Yes & Yes \\
  Major Economic Events & No & No & No & Yes & No & Yes \\
  \hline
  \textbf{Temperature} & & & & & & \\
  Spring Temp.  & 0.0030 & 0.0016 & 0.0004 & 0.0003 & 0.0020 & 0.0020 \\ 
                 & (0.0019) & (0.0033) & (0.0010) & (0.0010) & (0.0017) & (0.0017) \\ 
  Summer Temp.  & 0.0035 & 0.0058\textsuperscript{.} & -0.0017 & -0.0014 & 0.0019 & 0.0019 \\ 
                 & (0.0025) & (0.0030) & (0.0029) & (0.0029) & (0.0028) & (0.0028) \\ 
  Fall Temp.    & -0.0056\textsuperscript{.} & -0.0041\textsuperscript{.} & -0.0054 & -0.0054 & -0.0042 & -0.0042 \\ 
                 & (0.0028) & (0.0018) & (0.0034) & (0.0034) & (0.0033) & (0.0033) \\ 
  Winter Temp.  & -0.0084\textsuperscript{*} & -0.0082\textsuperscript{*} & -0.0019 & -0.0021 & -0.0055\textsuperscript{*} & -0.0055\textsuperscript{*} \\ 
                 & (0.0028) & (0.0029) & (0.0014) & (0.0015) & (0.0024) & (0.0024) \\ 
  (Spring Temp.)\verb|^|2 &       & -0.0013\textsuperscript{.} &       &       &       &       \\ 
                 &       & (0.0006) &       &       &       &       \\ 
  (Summer Temp.)\verb|^|2 &       & 0.0023 &       &       &       &       \\ 
                 &       & (0.0015) &       &       &       &       \\ 
  (Fall Temp.)\verb|^|2   &       & 0.0004 &       &       &       &       \\ 
                 &       & (0.0021) &       &       &       &       \\ 
  (Winter Temp.)\verb|^|2 &       & 0.0002 &       &       &       &       \\ 
                 &       & (0.0015) &       &       &       &       \\ 
  \hline
  \textbf{Precipitation} & & & & & & \\
  Spring Precip.  & 0.0003 & 0.0033 & -0.0003 & 0.0001 & 0.0013 & 0.0013 \\ 
                 & (0.0049) & (0.0069) & (0.0053) & (0.0053) & (0.0049) & (0.0049) \\ 
  Summer Precip.  & 0.0051 & 0.0115 & -0.0069\textsuperscript{*} & -0.0080\textsuperscript{*} & -0.0007 & -0.0007 \\ 
                 & (0.0063) & (0.0081) & (0.0029) & (0.0029) & (0.0043) & (0.0043) \\ 
  Fall Precip.    & 0.0035 & 0.0154 & 0.0091 & 0.0082 & 0.0069 & 0.0069 \\ 
                 & (0.0085) & (0.0105) & (0.0089) & (0.0089) & (0.0082) & (0.0082) \\ 
  Winter Precip.  & 0.0161\textsuperscript{.} & 0.0081 & 0.0001 & -0.0003 & 0.0078 & 0.0078 \\ 
                 & (0.0076) & (0.0059) & (0.0074) & (0.0076) & (0.0071) & (0.0071) \\ 
  (Spring Precip.)\verb|^|2 &       & -0.0373 &       &       &       &       \\ 
                 &       & (0.0228) &       &       &       &       \\ 
  (Summer Precip.)\verb|^|2 &       & -0.0064 &       &       &       &       \\ 
                 &       & (0.0229) &       &       &       &       \\ 
  (Fall Precip.)\verb|^|2  &       & -0.0728\textsuperscript{.} &       &       &       &       \\ 
                 &       & (0.0384) &       &       &       &       \\ 
  (Winter Precip.)\verb|^|2 &       & 0.0276 &       &       &       &       \\ 
                 &       & (0.0391) &       &       &       &       \\ 
  \hline
  \textbf{Model fit} & & & & & & \\
  AIC & -840.0440 & -837.1423 & -816.6137 & -814.6619 & -826.4337 & -824.4337 \\ 
  BIC & -716.6571 & -687.7792 & -735.4381 & -730.2393 & -742.0111 & -736.7640 \\ 
  \hline
\end{tabular}
\caption{Model fitting results. The coefficients are rounded to the nearest 4 digits. Significant codes $***, **, *$, and $.$ correspond to significance levels 0.001, 0.01, 0.05, and 0.1, respectively. The unit of measurement for precipitation shown in this table is percentage points. The robust clustered standard deviations are given in brackets.}
\label{tab:model_selection}
\end{table}

\begin{figure}[htbp!]
    \centering
    \includegraphics[width=0.8\textwidth]{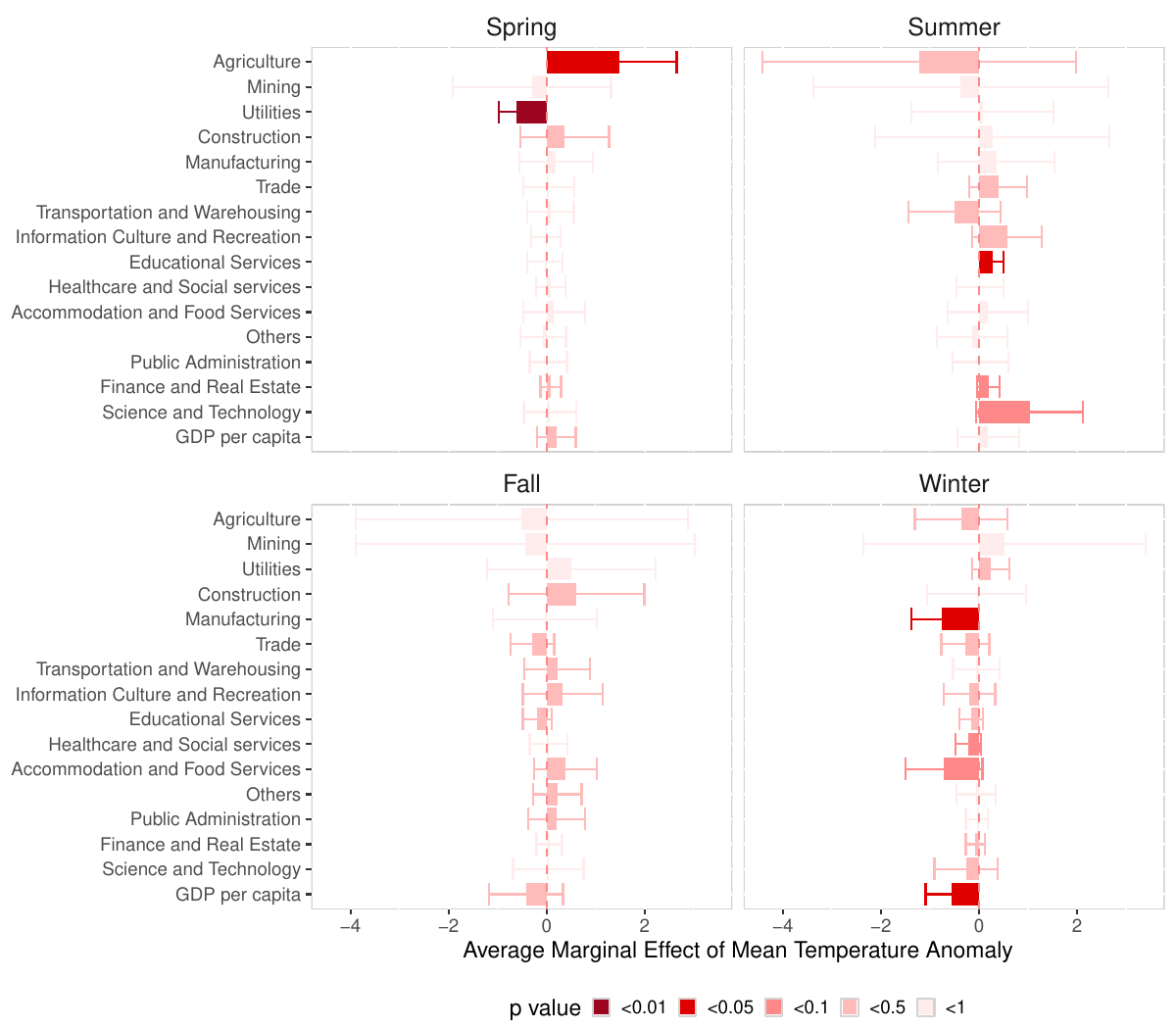}
    \includegraphics[width=0.8\textwidth]{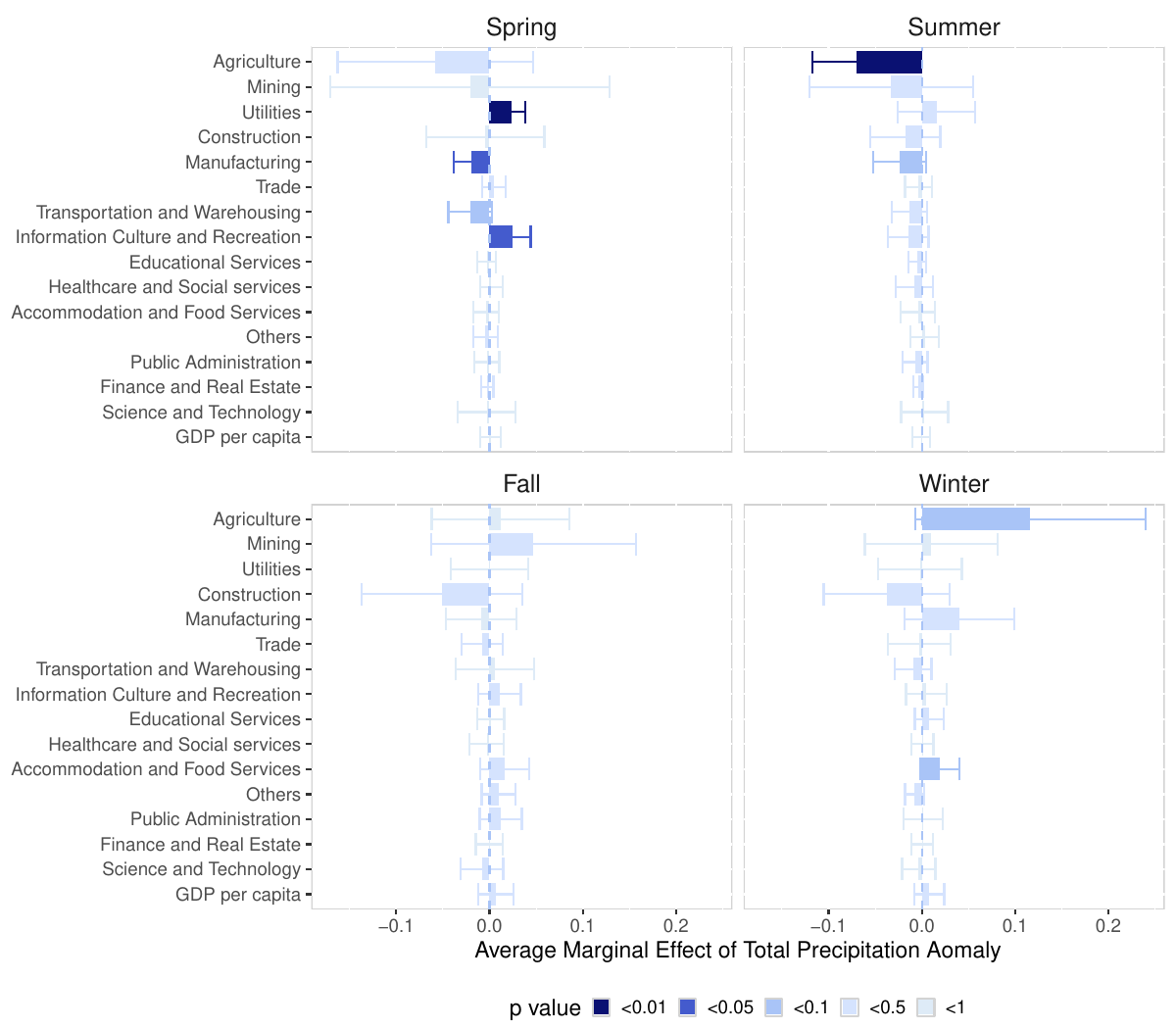}
    \caption{The average marginal effects of mean temperature anomaly and precipitation anomaly on industry GDP PCGR and their significance levels. Error bars denote the 95\% confidence intervals of the average marginal effects.}   
    \label{fig:industry-bars}
\end{figure}

Figure \ref{fig:industry-bars} presents the industry-specific results of climatic association with GDP PCGR. We observe that higher Winter Temperature anomalies are generally associated with declines in GDP PCGR across most industries. Thus, the overall effect of winter temperature anomalies on GDP growth likely reflects the combined influence of multiple sectors. In particular, warmer winters are associated with an average marginal effect (AME) of -0.76 on Manufacturing GDP PCGR (p-value $<0.05$). This effect may arise from reduced labor productivity, as higher temperatures may cause discomfort and reduce worker concentration, thereby lowering performance and efficiency \citep{kabore2023manufacturing, chen2023much}. In the Accommodation and Food Services industry, warmer winter conditions decrease demand for snow-dependent activities such as skiing and snowboarding due to diminished snow cover and unfavorable conditions \citep{elsasser2002climate, loomis1999estimated}. The heavy reliance on winter tourism may lead to lower GDP PCGR for the sector (AME: -0.71, p-value $<0.1$). Health Care and Social Services are also adversely affected (AME: -0.22, p-value $<0.1$), as warming is linked to an increased incidence and wider distribution of infectious diseases \citep{greer2008climate, carignan2019impact, ogden2019climate}, raising health burdens and system costs. A similar, though statistically insignificant, negative relationship between winter temperature anomalies and U.S. economic activities was reported by
\citet{colacito2019temperature}. The divergence in the magnitude and significance of effects likely reflects regional climatic and structural differences, such as Canada's shorter growing seasons, higher latitude exposure, and distinct industrial composition characterized by greater reliance on climate-sensitive sectors. Overall, our findings support and extend recent Canadian literature by demonstrating that, contrary to some prior expectations, warmer winters are associated with widespread adverse economic impacts rather than overall gains. 


In addition to the Winter Temperature anomaly, we find that Utilities GDP PCGR is negatively associated with the Spring temperature anomaly (AME: -0.62, p-value $<$ 0.01), which may reflect reduced energy demand during the transition between heating and cooling seasons, as temperature response functions are typically U-shaped \citep{rivers2020stretching}. 
In contrast, Agriculture GDP PCGR increases (AME: 1.48, p-value $<$ 0.05) as the Spring Temperature anomaly increases. This can be linked to enhanced growing conditions for agricultural crops during early stages, e.g., increased soil temperature \citep{gavito2001interactive, patil2010growth}, increased biomass allocation and nutrient uptake \citep{moorby1984effect}, and longer growing seasons. 


With respect to precipitation, we observe strong evidence that Agricultural GDP PCGR decreases with increasing Summer precipitation anomalies (AME: -0.075, p-value $<$ 0.05). This decline may be linked to excessive rainfall, which can reduce yields by causing waterlogging that disrupts crop development \citep{li2019excessive, beillouin2020impact, malik2002short}.
Moreover, a negative impact in Manufacturing GDP PCGR is observed as Spring precipitation anomaly increases (AME: -0.022, p-value $<$0.05). The mechanism is likely indirect: excessive precipitation can disrupt supply chains and damage infrastructure and transportation assets \citep{er2021modelling, liu2023global}, on which manufacturing heavily depends. Such precipitation shocks dampen growth across multiple sectors, with manufacturing particularly hindered \citep{kotz2022effect}.
We also observe strong evidence of GDP PCGR increase in Utilities (AME: 0.026, p-value $<$ 0.01) as Spring precipitation anomaly increases. In hydro-dominated provinces such as Quebec, British Columbia, and Manitoba \citep{statcan2024utilities}, additional spring rainfall and snowmelt may increase river flows and reservoir levels, boosting hydroelectric generation capacity and sectoral output \citep{wei2020effect}. 

While these findings are applicable nationwide, it is important to note that the results might exhibit a degree of heterogeneity when examined at the provincial level. This is because different geographical regions have distinct climate and economic patterns, as shown in Sections \ref{sector: econ} and \ref{sec: proj_weather}. Therefore, to further enhance our understanding of the possible climate change impacts, we calculate and investigate climate impact projections specific to provinces and industries in Section \ref{sec: projection}.

\FloatBarrier
\subsection{Projection of Climate Change Impact}\label{sec: projection}

\subsubsection{Climate Change Impact on Provincial GDP per capita}

\begin{figure}[h]
\centering
    \includegraphics[width=.7\textwidth]{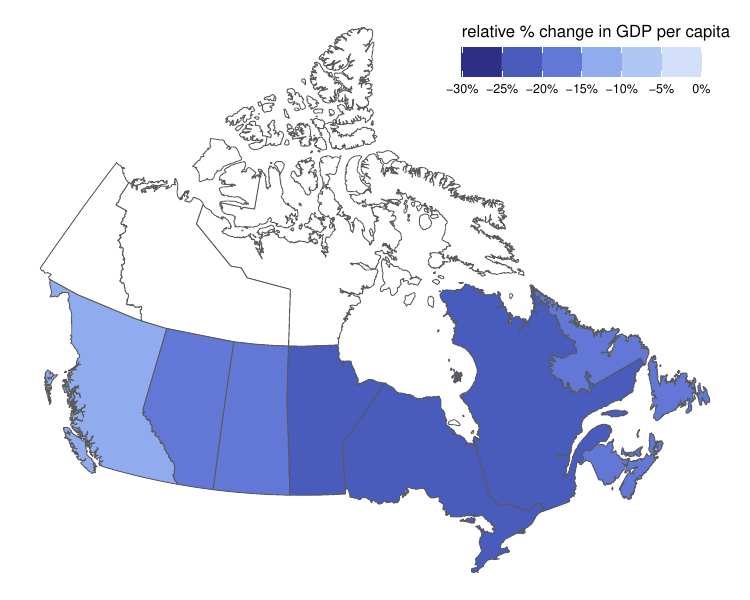}
    \caption{Expected impact of climate change on provincial GDP per capita by 2050 under RCP4.5.}
    \label{fig:proj_gdp}
\end{figure}

\begin{figure}[h]
    \centering
    \includegraphics[width=\textwidth]{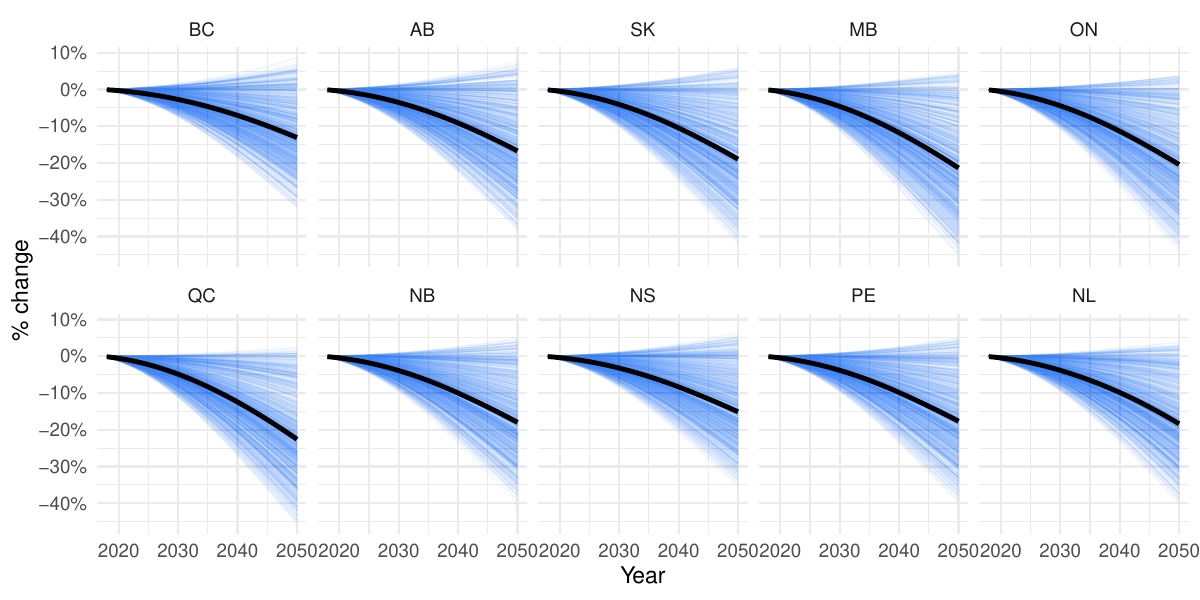}
    \caption{Projected trajectories of climate change impact on provincial GDP per capita under RCP4.5. The black line shows the projected trajectory of climate change impacts for each province under the estimated coefficients, while the blue lines represent the bootstrapped trajectories within the 95\% interval.}
    \label{fig:proj_ci}
\end{figure}

\begin{figure}[!h]
    \centering
    \includegraphics[width=\textwidth]{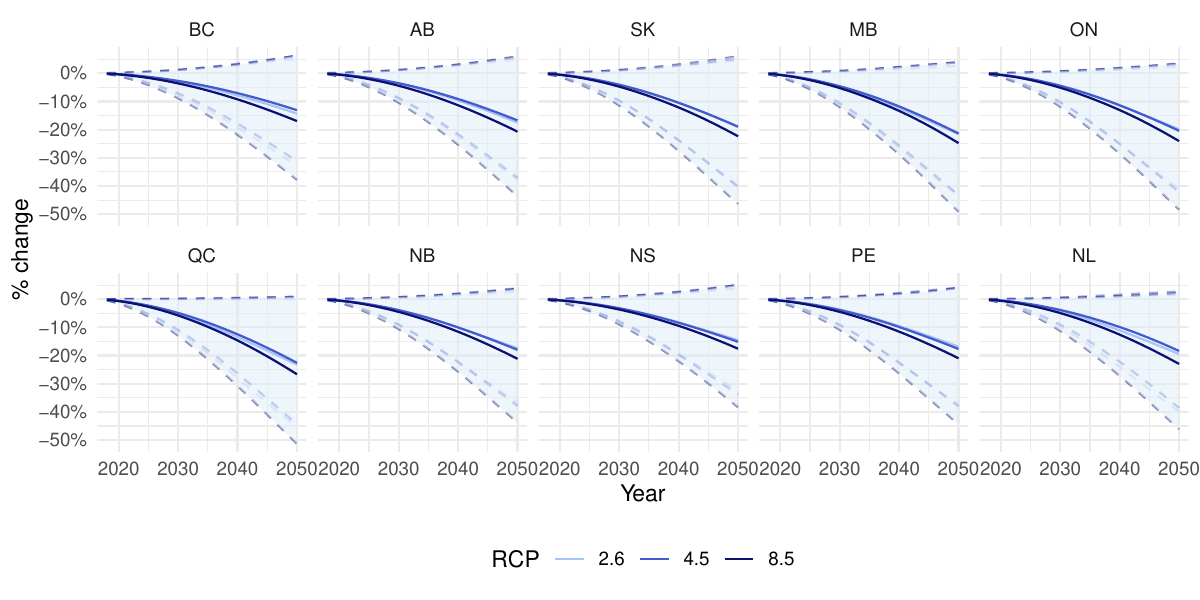}
    \caption{Projected trajectories of climate change impact on provincial GDP per capita under RCP2.6, RCP4.5, and RCP8.5. The dashed lines indicate the 2.5\% and 97.5\% bootstrapped quantiles of trajectories for each RCP scenario.}
    \label{fig:proj_ci_rcp}
\end{figure}

We provide our projection results for GDP per capita under RCP4.5 based on the estimated coefficients (Figure \ref{fig:proj_gdp}). The projections are expressed as relative changes in GDP, comparing outcomes under the climate-change scenario with those under a no-climate-change scenario, assuming no significant adaptation or structural economic changes (e.g., technological improvements or policy shifts). Therefore, our projection results can be thought of as representing the projected expectation over possible unexpected positive or negative climate or economic shocks. Overall, under RCP4.5, by 2050, all Canadian provinces are expected to experience a decline in GDP per capita by 13\% - 23\% compared to the without-climate-change scenario. This is the result of annual projected declines of sizes 0.4\% - 0.6\% accumulated over all projection years. This highlights that even if the impact is relatively minor in a given year, the consequences of climate change could be economically devastating when compounded over many years. 

Among all scenarios, Quebec, Manitoba, and Ontario are anticipated to experience larger declines, with projected relative decreases of 22.58\%, 21.37\%, and 20.41\%, respectively. On the other hand, British Columbia faces a smaller impact, with a projected relative decline of 13.11\%. The comparison drawn between central and coastal provinces reflects the geographical disparities in climate projections within Canada. 

We also report block-bootstrapped trajectories for GDP per capita under RCP4.5 (Figure \ref{fig:proj_ci}). These trajectories all cluster around the downward-sloping expectation, with sampling variability and forecast uncertainty compounding over time. While there is some heterogeneity in slope across provinces, the majority of resampled paths remain below zero by 2050. This suggests a high likelihood of adverse economic effects from climate change across Canada. Figure \ref{fig:proj_ci_rcp} further presents projected trajectories under three GHG emission scenarios (RCP2.6, RCP4.5, and RCP8.5), along with their 95\% prediction intervals. As emissions increase from RCP2.6 to RCP8.5, provinces face progressively larger declines in GDP per capita, accompanied by wider prediction intervals.

\subsubsection{Climate Change Impact by Industry}

\begin{figure}[h]
    \centering
    \includegraphics[width=.7\linewidth]{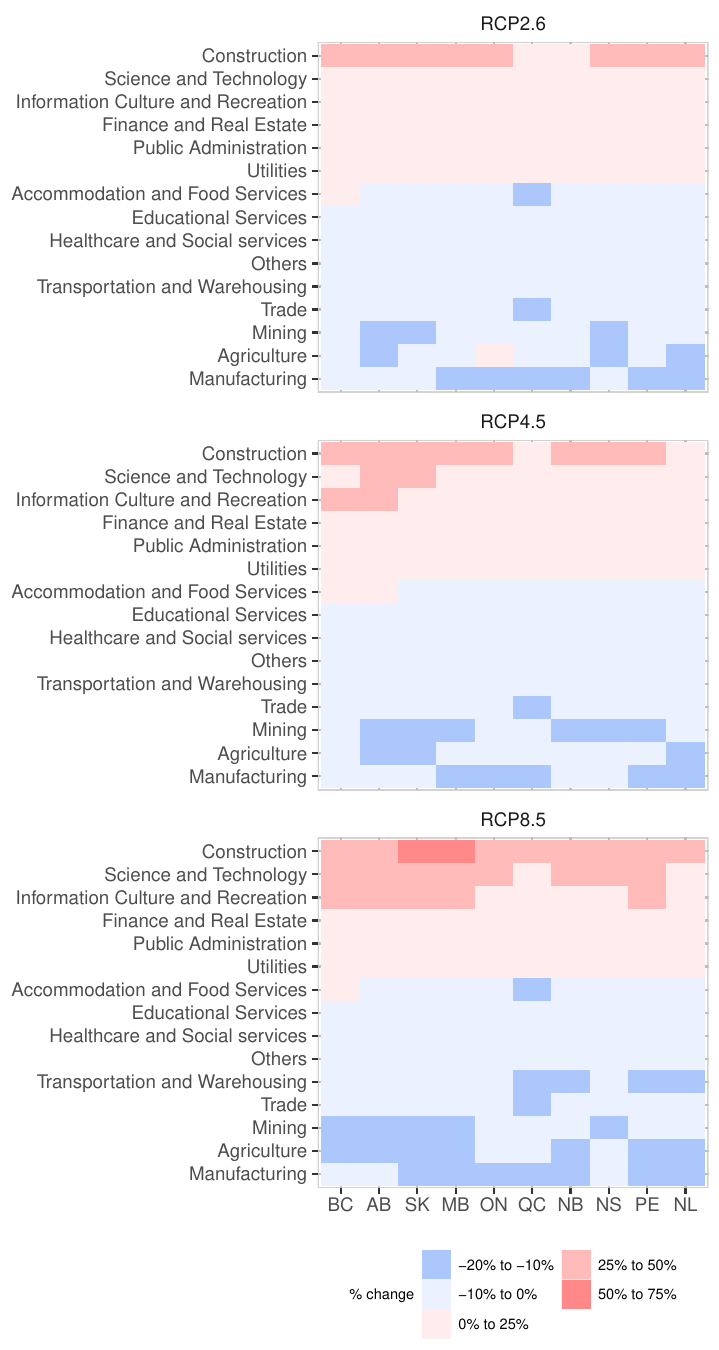}
    \caption{Projected impact of climate change by industry under RCP2.6, RCP4.5 and RCP8.5 by 2050.}
    \label{fig:proj_ind_rcp}
\end{figure}

Figure \ref{fig:proj_ind_rcp} presents the projected industry-specific impacts of climate change by 2050. Among all industries, Agriculture, Manufacturing, and Mining are projected to experience the largest adverse effects across different RCP scenarios. In contrast, certain sectors (e.g., Construction, Science and Technology, and Information, Culture and Recreation, etc.) are projected to benefit modestly from rising temperatures and precipitation anomalies. The projections also reveal spatial heterogeneity in impacts. For example, Agriculture in Alberta, Saskatchewan, and Newfoundland is expected to be more severely affected than in other provinces. This implies regional climatic conditions and indicates the uneven distribution of climate-related risks and opportunities across Canada. Moreover, as GHG emission pathways shift from RCP2.6 to RCP8.5, the magnitude of projected relative declines (and benefits) intensifies.

We examine the 95\% prediction interval of each industry using block bootstrap, with the results presented in Figure \ref{fig:proj_industry_ci} of Appendix~\ref{app:industry-bootstrap}. While sectors such as Agriculture, Manufacturing, and Mining exhibit the steepest expected relative declines, the bootstrapped trajectories span a wide range of outcomes, including positive, neutral, and negative relative growth. This wide dispersion suggests substantial sampling variability and implies that the impact of climate change on these industries contains considerable uncertainty. These patterns reflect the fact that when provinces are resampled in the bootstrap procedure, heterogeneous regional effects emerge: some provinces exhibit adverse impacts, whereas others display neutral or even favorable outcomes. Compared with the relatively narrow prediction intervals for aggregate GDP per capita, industry-level projections show much greater uncertainty. This is partly driven by variation in provincial industry shares and, in particular, by the heightened volatility of provinces with small shares in certain industries, especially in sectors such as Agriculture and Mining, where industry shares vary markedly across provinces. 

Nonetheless, there are several industries where the uncertainty surrounding projected impacts is relatively low. For example, the bootstrap trajectories for Finance and Real Estate, Information, Culture and Recreation, as well as Science and Technology, are more concentrated on the positive side of the distribution, indicating a stronger likelihood of growth under climate change scenarios. Unlike sectors such as Agriculture or Mining, the trajectories for these industries display greater alignment across provinces, suggesting that the observed positive trend is a broad national pattern. The comparatively narrow uncertainty bounds indicate that these sectors are likely to experience more robust growth and demonstrate greater resilience as climate patterns shift. 

\subsubsection{Sensitivity Analysis on Climate Change Impact}

We conduct sensitivity analyses of projections using alternative model specifications (outlined in Table \ref{tab:model_selection}), with results shown in Figures \ref{fig:proj_gdp_other} and \ref{fig:proj_ind_other} of Appendix~\ref{app:model-spec-proj}. This allows us to examine the consistency of our results under different modeling assumptions and methodologies. Several similar trends and characteristics can be observed across the projections, despite the differences in model specifications. At the provincial level, Quebec, Manitoba, and Ontario consistently stand out as the provinces projected to experience larger relative declines. This suggests these provinces exhibit a high level of vulnerability to climate change impacts. Conversely, British Columbia appears to be less affected across various model specifications and thus has more resilience in the face of projected changes related to climate. The recurrence of these patterns across various model specifications enhances our confidence in the projection results and underscores the difference in vulnerability against climate change among Canadian provinces. 

Industry-level sensitivity analyses reveal considerable uncertainty, though broad patterns persist. Across all non-quadratic models, Manufacturing, Agriculture, and Mining consistently experience the steepest relative declines (typically exceeding 10\%), followed by Trade, Transportation and Warehousing, and Others, which generally show moderate relative declines (0 - 10\%, occasionally higher). Healthcare and Social Services, Accommodation and Food Services, and Educational Services also show relative declines, but with less severity than the primary goods sectors. In contrast, Science and Technology, Information, Culture and Recreation, Finance and Real Estate, and Public Administration consistently display positive impacts. Some sectors, however, are highly sensitive to model specification: Construction alternates between sizable gains and losses, while Utilities show both mild positive and negative effects. These results suggest that certain sectors are more sensitive to modeling assumptions and therefore carry higher uncertainty in their projected responses to climate change.

Quadratic models produce more heterogeneity across both provinces and industries, reflecting increased variability in estimates. This added complexity leads to larger projection variances; nevertheless, model fit statistics caution against quadratic specifications, as they may introduce overfitting when applied to single-country data \citep{newell2021gdp}.

\subsubsection{Linking to Broader Evidence}

A substantial body of global literature highlights the adverse effects of rising temperatures and precipitation on productivity growth. For instance, \citet{colacito2019temperature} shows that rising temperatures could reduce U.S. economic growth by up to one-third over the next century. Long-term cross-country analysis by \citet{kahn2021} shows that persistent temperature and precipitation anomalies slow productivity and reduce global GDP per capita. Global macroeconomic analyses further show that persistent deviations of temperature from historical norms slow productivity growth and pose long-term risks to output and stability \citep{imf2019climate}. 

Recent official Canadian reports on the economic impacts of climate change align with our results to some extent. \cite{healthcosts2021} indicate that Quebec and Ontario are projected to experience the greatest losses in labour hours and to face the highest financial costs from reduced productivity due to heat, as well as the highest prevalence of Lyme disease. Manitoba is also projected to see substantial productivity losses, with British Columbia, New Brunswick, and Nova Scotia experiencing smaller but still noteworthy economic impacts~\citep{boyd2020costing}. Overall, climate change is expected to have uneven impacts on regional economic outcomes across Canada, reflecting variability in provincial economies and sectoral dependencies \citep{chartingcourse2020,netzero_future}. 

Evidence from global climate-economy studies further supports the sectoral patterns in our projections. For instance, the OECD report projects that Canadian crop yields will decline by 2050 \citep[Fig.~1.2]{oecd2015economic}, while U.S. evidence shows that heat strongly reduces agricultural output and labor productivity in Manufacturing, Mining, and other outdoor industries \citep{hsiang2017estimating, frbsf2024heat}.
Canadian sector-specific studies corroborate these global findings. The Canadian Climate Institute \citeyearpar{canadianclimateinstitute2024drought} documents severe recent impacts on Agriculture in the Prairies, including record declines in Saskatchewan’s crop production during the 2021 drought and unprecedented drought insurance payouts exceeding \$300 million to Alberta farmers in 2023. 
Microeconomic evidence further indicates that extreme temperatures reduce manufacturing output in Canada by lowering labor productivity and altering labor inputs \citep{kabore2023manufacturing}. 
The Mining sector is also projected to face substantial declines, reflecting multiple climate-related pressures. Canadian mining operations have historically been disrupted by extreme weather events \citep{pearce2009climate,pearce2011climate}. The sector further faces structural degradation and failures in transportation routes and critical mining infrastructure. Warmer winters also shorten the operating season of winter roads, which is critical for supplying remote mines. Beyond these physical risks, the industry's high carbon intensity, regulatory tightening, and significant upfront capital costs heighten its vulnerability to transition pressures. As a result, \citet{chartingcourse2020} projects that the sector will be particularly affected by lower revenues, profits, and employment levels. 


On the contrary, Finance and Real Estate, Science and Technology, and Information, Culture and Recreation consistently exhibit positive impacts. It is plausible that these sectors may benefit under projected climate conditions. Global modeling by the OECD projects that the Canadian Other Services sector shows positive value-added changes and an increasing share in 2060 under climate change scenarios \citep[Fig.~2.11]{oecd2015economic}. Unlike sectors tied to physical production, these sectors may face indirect benefits from climate-driven changes. For instance, climate change generates growing demand for clean energy, climate monitoring, and green technologies. Empirical evidence across OECD countries further shows that technological innovation, renewable energy adoption, and environmental innovation significantly contribute to green growth \citep{qamruzzaman2024green}. More broadly, climate action itself may catalyze new economic opportunities by channeling investment flows into clean energy and related industries \citep{oecd2025climate}.

\FloatBarrier
\section{Conclusions, Limitations, and Policy Recommendations}

\subsection{Conclusions}

In this study, we have explored how seasonal climate variables influence the GDP PCGR for each Canadian province and industry. Through the use of various linear mixed-effects models, we are able to examine the impact of seasonal climate anomalies while accounting for potential unobserved factors that might influence the GDP PCGR across provinces and over time. 
Our model results consistently indicate evidence supporting the negative effect of rising Winter temperature anomalies on GDP growth. Furthermore, we extended our analysis to individual industries and found that certain sectors, such as Manufacturing, are significantly associated with Winter temperature anomalies.

To assess the potential impact of climate change on GDP per capita across Canadian provinces and industries, we calculated projections of climate change impact considering different RCP scenarios up to the year 2050. Across all RCP scenarios, relative declines are consistently observed in every province in Canada. The largest projected relative decreases occur in Quebec, Manitoba, and Ontario, reflecting both the rapid pace of regional climate change and these provinces' greater reliance on climate-sensitive sectors such as Manufacturing. In contrast, British Columbia is projected to be less affected, which may be partly attributable to its relatively smaller temperature increases and limited exposure to climate-sensitive sectors.

Our industry-level projections indicate that the impacts of climate change vary substantially across sectors. Certain sectors, such as Finance and Real Estate, Science and Technology, and Information, Culture and Recreation might be better positioned to thrive in the projected climate conditions, while others, such as Agriculture, Manufacturing, and Mining, could face more challenges and economic declines. Our projection results under different modeling assumptions found shared trends and characteristics, which strengthen the robustness of evidence supporting our conclusions. The findings observed in our provincial and industry projections may provide valuable insights into the differential impacts of climate change across various industries, which possibly inform policy-making and strategic planning efforts.

\subsection{Limitations of Our Study}

It is important to acknowledge the limitations of our study. First, our projections were based on extrapolation beyond the support of the existing model, which may be sensitive to abrupt changes in future climate and economic patterns, including natural disasters or climate actions. Second, our model only incorporated two weather variables, which may not capture the full complexity of the relationship between climate and GDP growth. Third, to reduce the confounding effects, we have incorporated key economic indices, including commodity prices, unemployment rates, interest rates, world GDP, and major economic events. While this provides broad macroeconomic coverage, certain factors (e.g., technological change) remain unobserved and could still introduce bias into the estimates.

\subsection{Policy Recommendations}

Our projections indicate a high likelihood of adverse economic impacts from climate change across Canada. Although our analysis extends to 2050, the economic consequences are likely to intensify beyond mid-century, further eroding Canada's long-term economic stability. Therefore, achieving Canada's Net-Zero targets and strengthening adaptation initiatives are critical to limit long-term damages, reduce economic volatility, and align with global climate commitments \citep{netzero, netzero_future}.

The differences in projected economic outcomes between RCP2.6 and RCP8.5 highlight the importance of strong mitigation. Under a low-emissions pathway (RCP2.6), Canadian provinces experience smaller relative GDP losses and narrower uncertainty intervals, whereas the high-emissions pathway (RCP8.5) is associated with substantially greater declines across both aggregate and sectoral levels. This divergence illustrates the significant economic value of emissions reduction, as mitigation directly reduces the scale of damages that future adaptation efforts would need to address. 

At the same time, our results reveal a heterogeneous climate change impact across provinces, reflecting differences in regional exposure and economic structure. Despite Canada's nationwide installation of the Net-Zero by 2050 initiative, it is vital to account for the disparities in climate change impacts among different provinces. This is particularly critical for vulnerable provinces, as they may need extra attention, assistance, and resources to withstand the potential climatic challenges. In addition, not all provinces have fully embraced the long-term goal for GHG emissions \citep{climatechageaudit, pancandian}. For instance, Saskatchewan and Manitoba have not fully established their own carbon pricing mechanisms. Recognizing these differences can help provincial leaders, policymakers, and communities adapt to region-specific challenges, address vulnerabilities, and leverage local strengths through climate-smart strategies. These findings underscore the urgent need for coordinated action across federal and provincial levels, supported by regular monitoring and periodic reassessment of climate-economic impacts to ensure timely and effective policy guidance.

The projection differences among industries also emphasize the need for the government to adopt a tailored approach to policy-making, as sectors possess different levels of intricacies and vulnerabilities in the face of climate change. 
By acknowledging the differences among industries, governments can strategically allocate resources and efforts to address the challenges these industries face. 
For Manufacturing, policies should prioritize investment in sustainable production technologies and infrastructure upgrades to minimize supply chain disruptions, alongside region-specific measures such as enhanced cooling infrastructure and heat preparedness in Quebec and Ontario.
For Agriculture, this may involve diversifying cropping systems, improving water management, and investing in drought-resistant crop varieties, particularly in the Prairie and Maritime provinces. 
For Mining, strengthening infrastructure resilience in vulnerable regions such as Alberta, Saskatchewan, and Newfoundland will be critical to sustaining operations under increasing climate stress. It is also important that mining firms adopt proactive transition strategies by investing in low-carbon technologies, improving energy efficiency, and integrating climate risk assessments into long-term planning.

Since Manufacturing and other resource-based industries are labor-intensive, sectoral declines risk triggering job losses and exacerbating unemployment. In addition to mitigation and adaptation, a ``just transition" approach is therefore essential to ensure fairness in the shift toward a low-carbon economy \citep{ilo2015just, unfccc2021just, cicc2021employment}. Such an approach would address potential employment dislocations in carbon-intensive industries through training and re-skilling programs, while also mitigating regional inequalities in provinces more dependent on vulnerable sectors. By integrating ``just transition" policies into sector-specific strategies, governments can both strengthen resilience and promote equity and social cohesion in Canada's climate response.

The uneven sectoral impacts of climate change carry important implications for sustainable finance. Resilient sectors such as Finance \& Real Estate, Science \& Technology, and Information, Culture \& Recreation are better positioned to attract investment flows, benefit from favorable financing terms, and expand their market share under projected climate conditions. In contrast, climate-sensitive sectors, including Agriculture, Manufacturing, and Mining, face rising transition and physical risks that may undermine revenues, reduce asset values, and increase borrowing costs. For financial institutions, this divergence highlights the need to reassess portfolio exposures, integrate climate scenarios into financial monitoring, and develop stress-testing frameworks that capture climate-related risks \citep{battiston2017climate, ngfs2019call}. Beyond risk management, it also points to the strategic opportunity of reallocating capital toward resilient, sustainable, and low-carbon industries that are better positioned to drive long-term economic growth \citep{oecd2025climate}. This transition in financing priorities is critical not only to safeguard portfolio stability but also to align investment decisions with broader climate goals and ensure long-term economic resilience.



\section*{Acknowledgements}

The authors would like to thank the Statistical Society of Canada (SSC) for hosting the case study competition that helped inspire the analysis presented in this paper. This work was partially supported by the Natural Sciences and Engineering Research Council of Canada under Discovery Grant RGPIN-2019-04771.

\section*{Disclosure Statement}

The authors report there are no competing interests to declare.

\section*{Data Availability Statement}

The data analyzed in this paper are publicly available from Statistics Canada (URL: https://www150.statcan.gc.ca/t1/tbl1/en/tv.action?pid=3610040001, https://www150.statcan.gc.ca/t1/tbl1/en/tv.action?pid=1710000501 and https://www150.statcan.gc.ca/t1/tbl1/en/tv.action?pid=1410032701), World Bank (https://data.worldbank.org/indicator/NY.GDP.MKTP.CD and https://www.worldbank.org/en/research/commodity-markets), and Government of Canada
(URL: \href{https://climate-change.canada.ca/climate-data/#/monthly-climate-summaries}{https://climate-change.canada.ca/climate-data/\#/monthly-climate-summaries}, https://climate-scenarios.canada.ca/?page=cmip6-scenarios, and https://open.canada.ca/data/en/dataset/d5ffb2fb-3607-4bda-8f74-2f7b264a9a85). 

\FloatBarrier
\newpage
\bibliographystyle{chicago}
\bibliography{paper}

\FloatBarrier
\newpage
\appendix
\section*{Appendix}

\setcounter{section}{0}
\renewcommand{\thesection}{\Alph{section}}

\section{List of Selected High-Impact Economic Events} \label{app:econ-events}

\renewcommand{\arraystretch}{1.05}
\begin{center}
\footnotesize
\begin{longtable}{|p{7cm}|c|c|p{5cm}|}
\hline
\textbf{Event} & \textbf{Year} & \textbf{Month} & \textbf{Impacted Province(s)} \\
\hline
\hline
\endfirsthead
\hline
\textbf{Event} & \textbf{Year} & \textbf{Month} & \textbf{Impacted Province(s)} \\
\hline
\endhead
1. Asian Financial Crisis  & 1997 & 7 & All\footnote{All: NL, PE, NS, NB, QC, ON, MB, SK, AB, BC.} \\ \hline
2. Russian Financial Crisis & 1998 & 8 & All \\ \hline
3. Swissair Flight 111 crash & 1998 & 9 & NS \\ \hline
4. Tech Bubble Economic Boom & 2000 & 3 & All \\ \hline
5. Walkerton E. coli outbreak & 2000 & 5 & ON \\ \hline
6. 2001 September 11 Attacks & 2001 & 9 & All \\ \hline
7. West Nile Virus Outbreak (First human cases in Canada) & 2002 & 7 & ON, QC, NS, MB, SK, AB, BC \\ \hline
8. 2003 SARS Outbreak & 2003 & 3 & ON, BC \\ \hline
9. Mad Cow Disease (BSE) Outbreak & 2003 & 5 & AB, SK, MB, ON \\ \hline
10. 2003 Northeast Blackout & 2003 & 8 & ON \\ \hline
11. Alberta Specified Gas Emitters Regulation Begins & 2007 & 7 & AB \\ \hline
12. BC Carbon Tax Announced & 2008 & 2 & BC \\ \hline
13. Toronto Land Transfer Tax Introduced & 2008 & 2 & ON \\ \hline
14. BC Carbon Tax Implemented & 2008 & 7 & BC \\ \hline
15. Global financial crisis (Lehman Brothers collapse) & 2008 & 9 & All \\ \hline
16. H1N1 influenza outbreak & 2009 & 4 & All \\ \hline
17. Vancouver Winter Olympics & 2010 & 2 & BC \\ \hline
18. Toronto G20 protests & 2010 & 6 & ON \\ \hline
19. Vancouver Stanley Cup Riot & 2011 & 6 & BC \\ \hline
20. Quebec student protests ("Maple Spring") & 2012 & 2 & QC \\ \hline
21. Quebec Cap-and-Trade Law Passed & 2012 & 12 & QC \\ \hline
22. Lac-Megantic train derailment & 2013 & 7 & QC \\ \hline
23. Quebec and California Cap-and-Trade Linkage & 2014 & 1 & QC \\ \hline
24. Global oil price shock (Start of major decline) & 2014 & 6 & NL, SK, AB \\ \hline
25. Oil Price Drop Following OPEC Decision & 2014 & 11 & AB, SK \\ \hline
26. Alberta oil sands royalties review report released & 2016 & 1 & AB \\ \hline
27. BC Property Transfer Tax Increased (Luxury Thresholds) & 2016 & 2 & BC \\ \hline
28. Alberta Broad Carbon Levy Introduced (Climate Leadership Plan) & 2016 & 5 & AB \\ \hline
29. Vancouver Foreign Buyers Tax Introduced (15\%) & 2016 & 8 & BC \\ \hline
30. Canada-EU Comprehensive Economic and Trade Agreement (CETA) Signed & 2016 & 10 & All \\ \hline
31. Vancouver Empty Homes Tax (Vacancy Tax) Announced & 2016 & 11 & BC \\ \hline
32. Toronto Municipal Land Transfer Tax Increased & 2017 & 1 & ON \\ \hline
33. Interprovincial Free Trade Agreement (CFTA) Announced & 2017 & 4 & All \\ \hline
34. Ontario Non-Resident Speculation Tax (15\%) Announced & 2017 & 4 & ON \\ \hline
35. Softwood Lumber Tariff (US Imposition) & 2017 & 4 & BC, QC \\ \hline
36. Manitoba Green Levy Announced & 2017 & 10 & MB \\ \hline
37. Ontario, Quebec Cap-and-Trade Implemented & 2017 & 9 & ON, QC \\ \hline
38. CETA Implementation (Provisional) & 2017 & 9 & All \\ \hline
\end{longtable}
\end{center}
\newpage

\section{Sensitivity Analysis of Model Specification} \label{app:sensitive-model-spec}

In addition to the six models reported in Table~\ref{tab:model_selection} of the main paper, we fitted alternative versions of the general model in Equation~\eqref{eq:main_model} to assess the sensitivity of our results. In particular, Model (1S) extends the two-way fixed effects framework by incorporating random effects of major economic events to capture province-time heterogeneity. Model (2S) includes interaction terms between seasonal temperature and precipitation, while Model (3S) adds province-specific time trends. Model (4S) allows for province-specific random slopes of climate effects. Finally, Models (5S) and (6S) mirror the specifications of Models (2S) and (3S), respectively, but replace fixed year effects with random year effects and incorporate economic indices. The results are shown in Table~\ref{tab:model-sensitivity} below. We can see that Winter temperature has significant negative effects across all model specifications, while evidence for other climate factors is
less clear. These patterns align with the findings reported in the main paper.

\begin{table}[h]
\scriptsize
\renewcommand{\arraystretch}{1.1}
\centering
\begin{tabular}{lccccccc}
  \hline
  & (1S) & (2S) & (3S) & (4S) & (5S) & (6S) \\ 
  \hline
  \textbf{Model specification} \\
  Year Effect & Fixed & Fixed & Fixed & Fixed & Random & Random \\
  Economic Indices & No & No & No & No & Yes & Yes \\
  Major Economic Events & Yes & No & No & No & No & No \\
  \hline
  \textbf{Temperature} \\
  Spring Temp. & 0.003  & 0.0022  & 0.0019  & 0.0031. & 0.0018  & 0.0002  \\ 
    & (0.0019) & (0.0016) & (0.0011) & (0.0017) & (0.0016) & (0.0014) \\ 
  Summer Temp. & 0.0035  & 0.0045. & 0.0012  & 0.0045  & 0.0024  & -0.0016  \\ 
    & (0.0025) & (0.0022) & (0.0038) & (0.0031) & (0.0027) & (0.004) \\ 
  Fall Temp. & -0.0056. & -0.0054. & -0.0063. & -0.0023  & -0.0041  & -0.004  \\ 
    & (0.0028) & (0.0025) & (0.003) & (0.0028) & (0.0033) & (0.0032) \\ 
  Winter Temp. & -0.0084* & -0.0087* & -0.008** & -0.0085** & -0.0056* & -0.0039. \\ 
    & (0.0028) & (0.0029) & (0.0025) & (0.0023) & (0.0024) & (0.002) \\ 
  \hline
  \textbf{Precipitation} \\
  Spring Precip. & 0.0003  & 0.0032  & -0.0016  & 0.0009  & 0.0021  & 0.0002  \\ 
    & (0.0049) & (0.0054) & (0.0054) & (0.0057) & (0.0053) & (0.004) \\ 
  Summer Precip. & 0.0051  & 0.0082  & 0.0067  & 0.0096  & 0.0017  & 0.0003  \\ 
    & (0.0063) & (0.0073) & (0.0072) & (0.0109) & (0.0049) & (0.0046) \\ 
  Fall Precip. & 0.0035  & 0.0046  & 0.0016  & 0.0122  & 0.008  & 0.0044  \\ 
    & (0.0085) & (0.0079) & (0.0106) & (0.0174) & (0.0085) & (0.0099) \\ 
  Winter Precip. & 0.0161. & 0.0136  & 0.0125  & 0.0184* & 0.0066  & 0.0021  \\ 
    & (0.0076) & (0.0085) & (0.0101) & (0.0071) & (0.0081) & (0.0105) \\ 
  \hline
  \textbf{Others} \\
  Spring Temp. $\times$ Spring Precip. &   & -0.0057  &  &  & -0.0024  &  \\ 
    &  & (0.0037) &  &  & (0.0032) &  \\ 
  Summer Temp. $\times$ Summer Precip. &  & -0.0125  &  &  & -0.0025  &  \\ 
    &  & (0.0109) &  &  & (0.0105) &  \\ 
  Fall Temp. $\times$ Fall Precip. &  & -0.0078  &  &  & -0.0065  &  \\ 
    &  & (0.008) &  &  & (0.0084) &  \\ 
  Winter Temp. $\times$ Winter Precip. &  & -0.0015  &  &  & -0.0014  &  \\ 
    &  & (0.0033) &  &  & (0.0038) &  \\ 
  \hline 
  \textbf{Model fit} \\
  AIC & -838.0440 & -835.3294 & -844.2899 & -795.7559 & -819.4359 & -836.0570 \\ 
  BIC & -711.4101 & -698.9544 & -691.6798 & -555.4761 & -722.0252 & -719.1642 \\ 
   \hline
\end{tabular}
\caption{Model fitting sensitivity to model specification. The coefficients are rounded to the nearest 4 digits. Significant codes $***, **, *$, and $.$ correspond to significance levels 0.001, 0.01, 0.05, and 0.1, respectively. The unit of measurement for precipitation shown in this table is percentage points. The robust clustered standard deviations are given in brackets.}
\label{tab:model-sensitivity}
\end{table}

\vspace{-10mm}
\section{Industry-Specific Projection Results with Uncertainty} \label{app:industry-bootstrap}

\FloatBarrier
\begin{figure}[htbp!]
\captionsetup[subfigure]{labelformat=empty}
    \begin{subfigure}[b]{\linewidth}
         \centering
             \includegraphics[width=.9\linewidth]{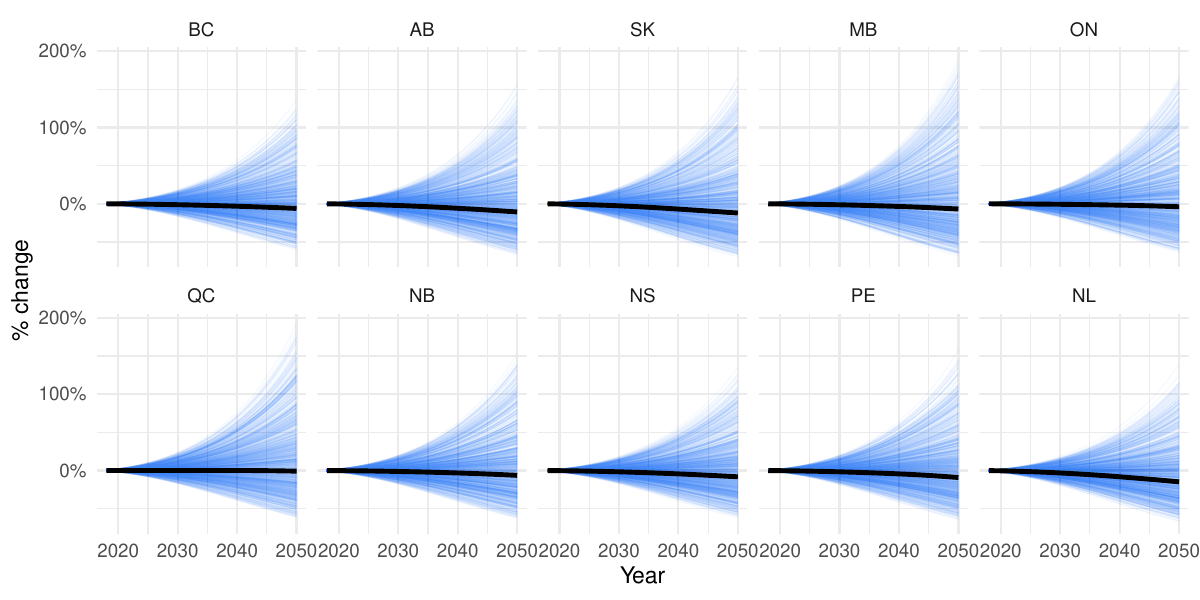}
         \caption{(a) Agriculture}
    \end{subfigure} 
    \begin{subfigure}[b]{\linewidth}
         \centering
         \includegraphics[width=.9\linewidth]{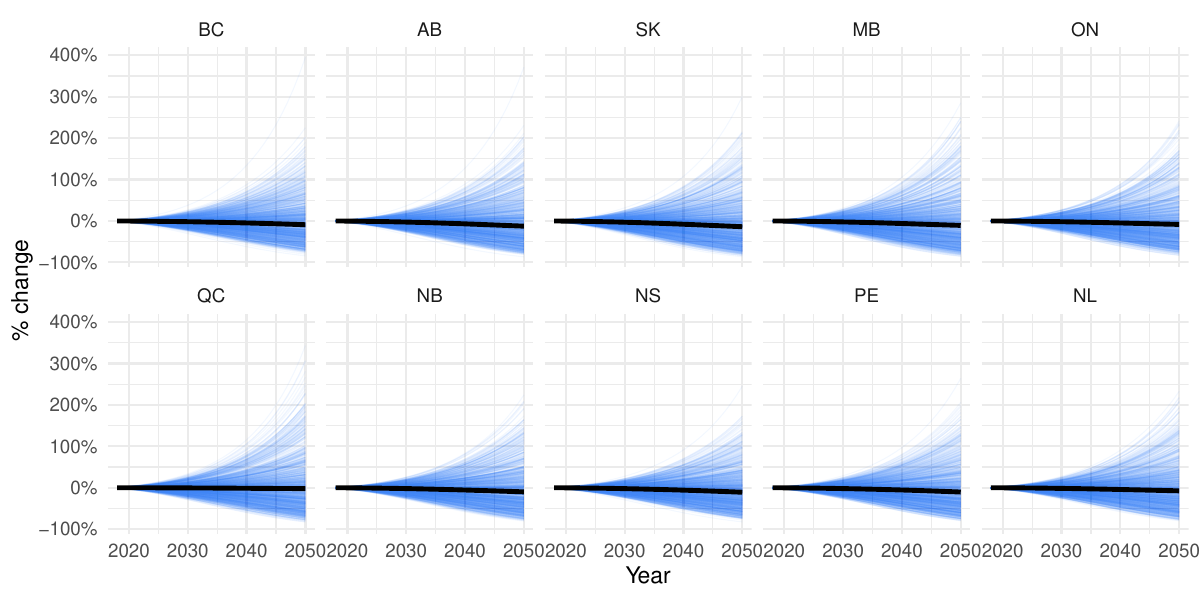}
         \caption{(b) Mining}
    \end{subfigure} 
    \begin{subfigure}[b]{\linewidth}
         \centering
         \includegraphics[width=.9\linewidth]{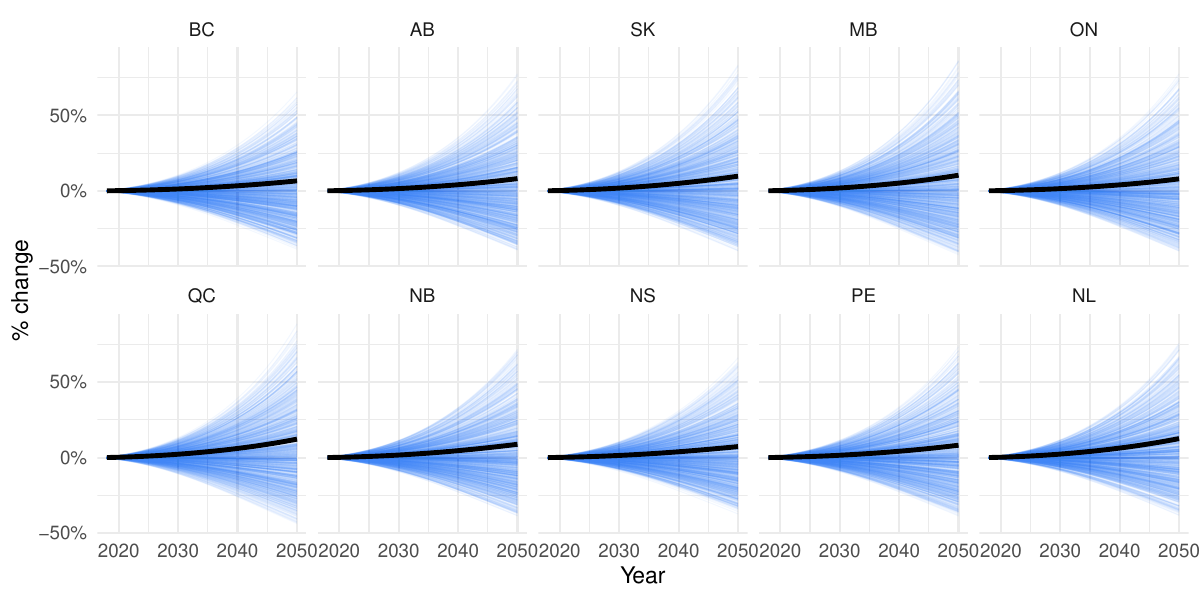}
         \caption{(c) Utilities}
    \end{subfigure} 
    \caption{Projected impact of climate change on GDP per capita across provinces and industries under RCP4.5 with uncertainty calculated using province-based block bootstrap.} \label{fig:proj_industry_ci}
\end{figure}

\begin{figure}[htbp]\ContinuedFloat
\captionsetup[subfigure]{labelformat=empty}
    \begin{subfigure}[b]{\linewidth}
         \centering
             \includegraphics[width=.9\linewidth]{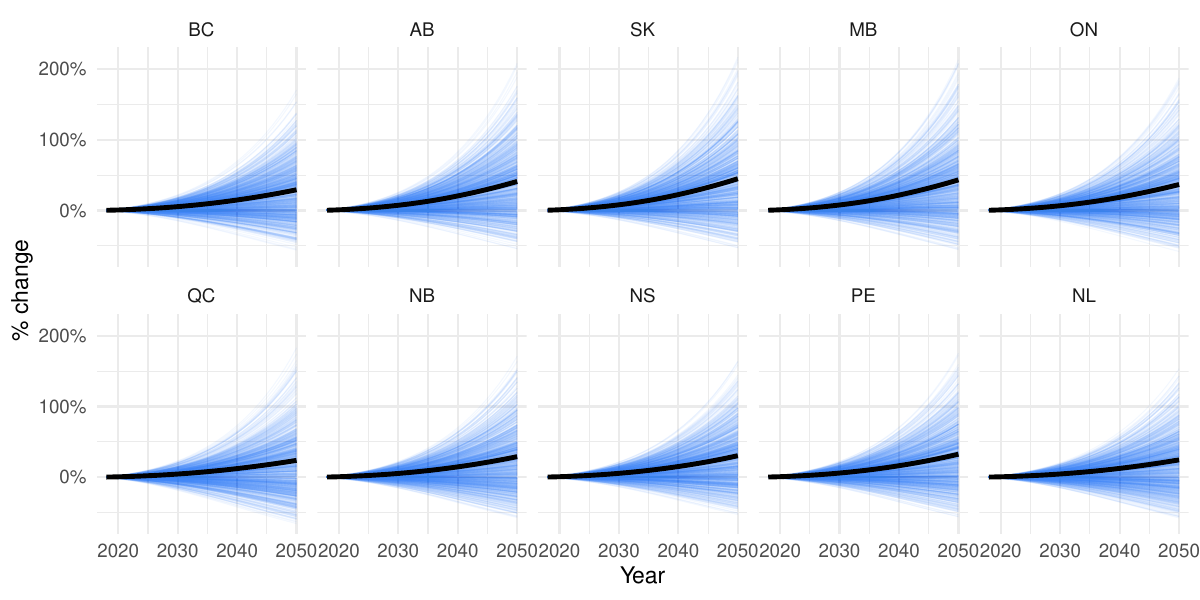}
         \caption{(d) Construction}
    \end{subfigure} 
    \begin{subfigure}[b]{\linewidth}
         \centering
         \includegraphics[width=.9\linewidth]{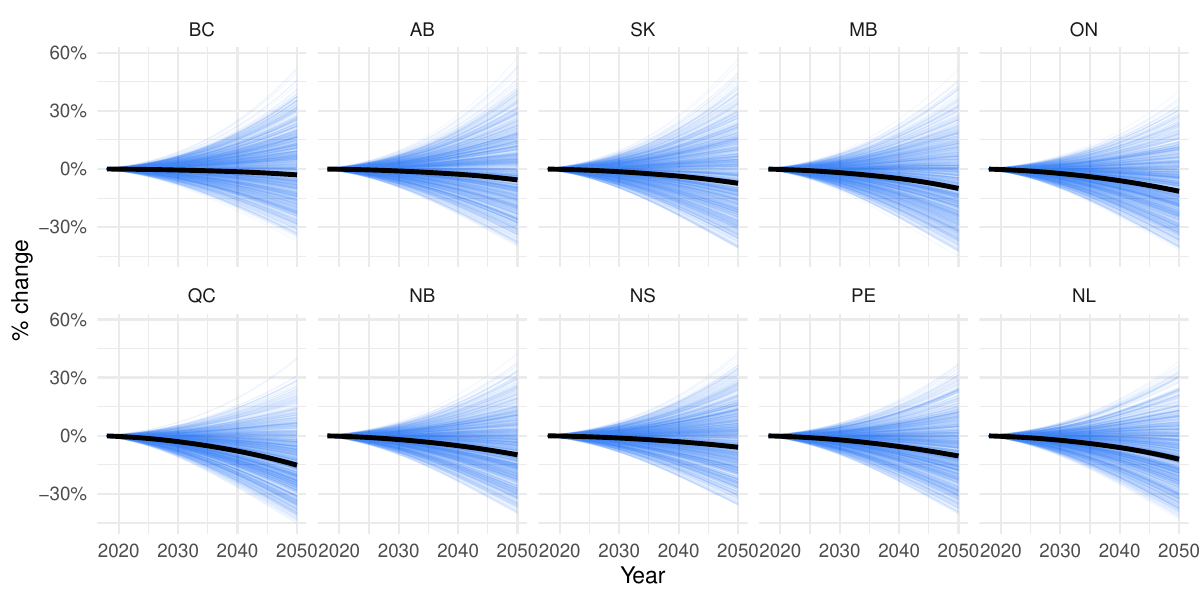}
         \caption{(e) Manufacturing}
    \end{subfigure} 
    \begin{subfigure}[b]{\linewidth}
         \centering
         \includegraphics[width=.9\linewidth]{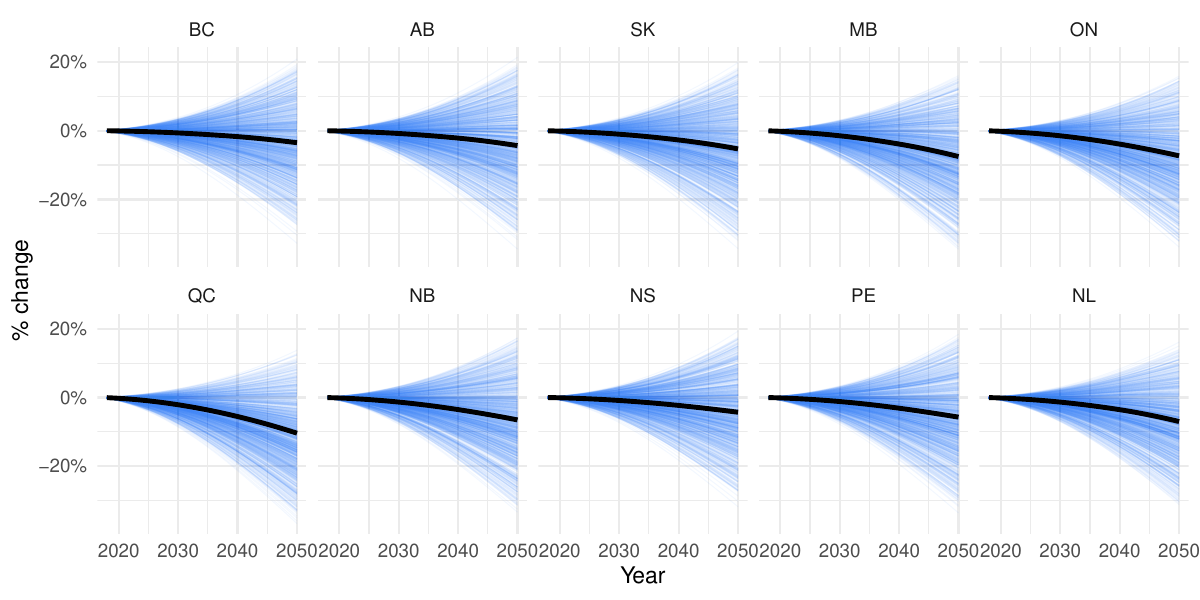}
         \caption{(f) Trade}
    \end{subfigure} 
    \caption{Projected impact of climate change on GDP per capita across provinces and industries under RCP4.5 with uncertainty calculated using province-based block bootstrap.} 
\end{figure}

\begin{figure}[htbp]\ContinuedFloat
\captionsetup[subfigure]{labelformat=empty}
    \begin{subfigure}[b]{\linewidth}
         \centering
             \includegraphics[width=.9\linewidth]{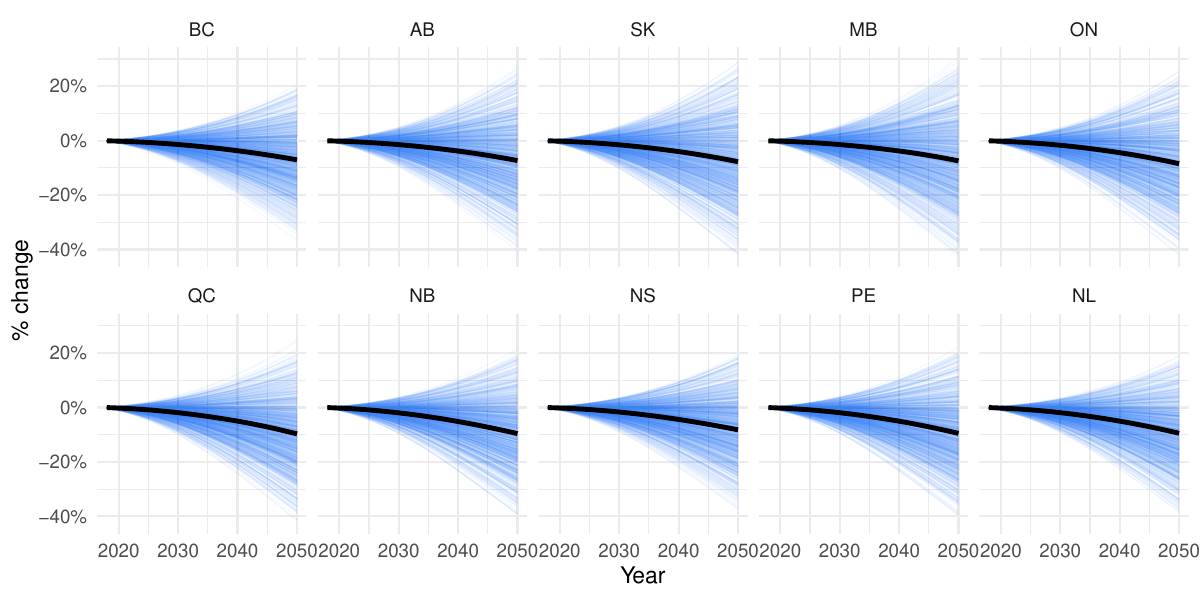}
         \caption{(g) Transportation and Warehousing}
    \end{subfigure} 
    \begin{subfigure}[b]{\linewidth}
         \centering
         \includegraphics[width=.9\linewidth]{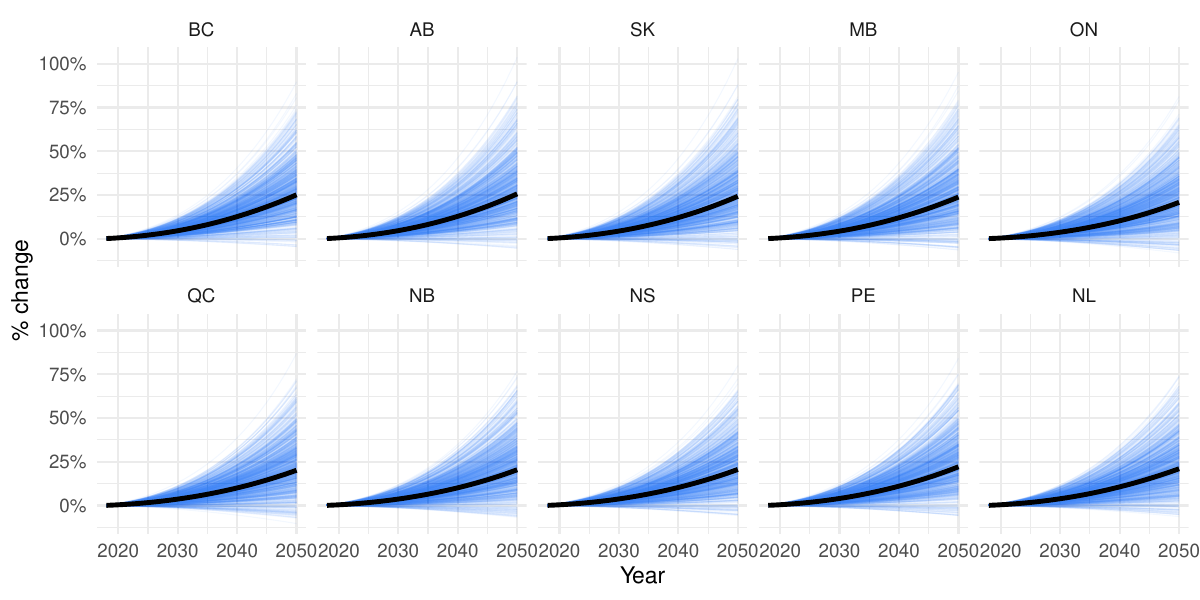}
         \caption{(h) Information, Culture and Recreation}
    \end{subfigure} 
    \begin{subfigure}[b]{\linewidth}
         \centering
         \includegraphics[width=.9\linewidth]{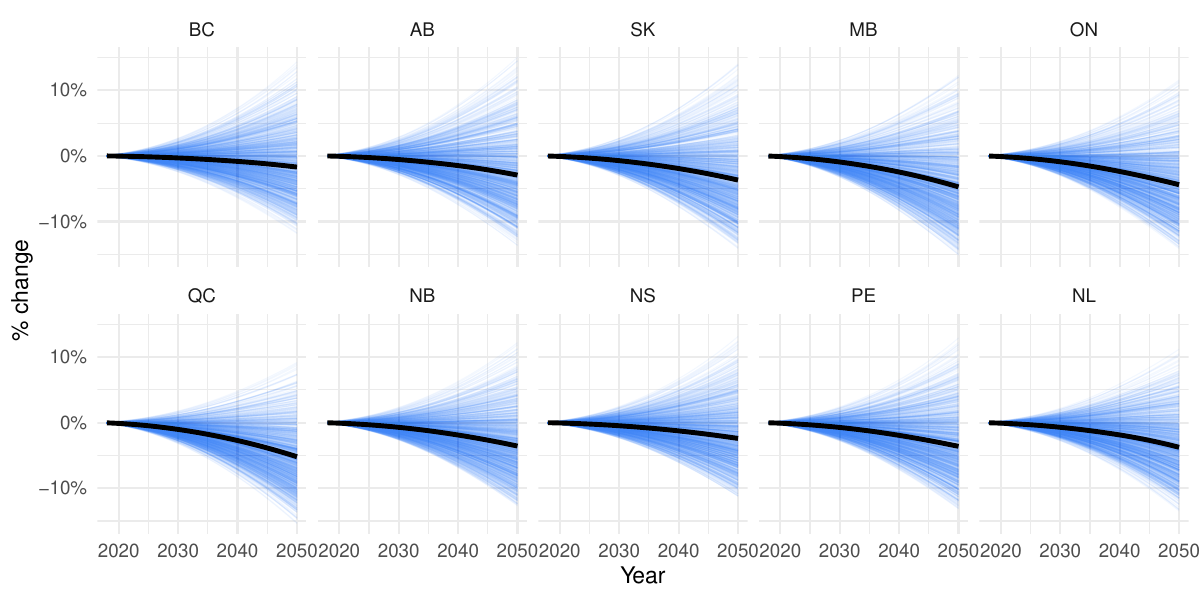}
         \caption{(i) Educational Services}
    \end{subfigure} 
    \caption{Projected impact of climate change on GDP per capita across provinces and industries under RCP4.5 with uncertainty calculated using province-based block bootstrap.} 
\end{figure}

\begin{figure}[htbp]\ContinuedFloat
\captionsetup[subfigure]{labelformat=empty}
    \begin{subfigure}[b]{\linewidth}
         \centering
             \includegraphics[width=.9\linewidth]{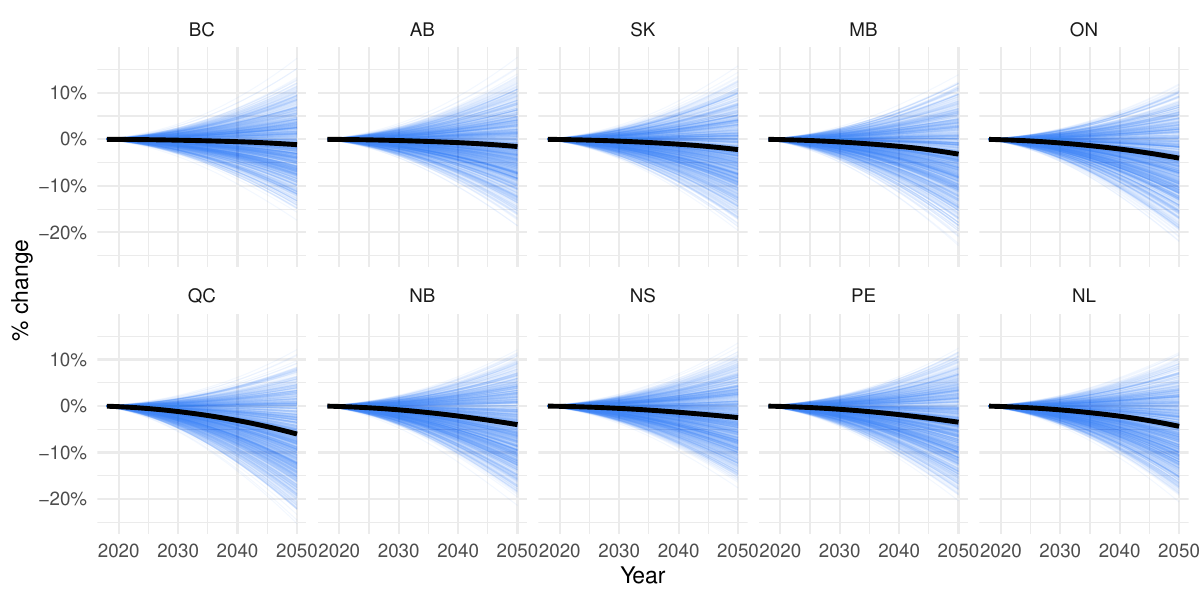}
         \caption{(j) Healthcare and Social services}
    \end{subfigure} 
    \begin{subfigure}[b]{\linewidth}
         \centering
         \includegraphics[width=.9\linewidth]{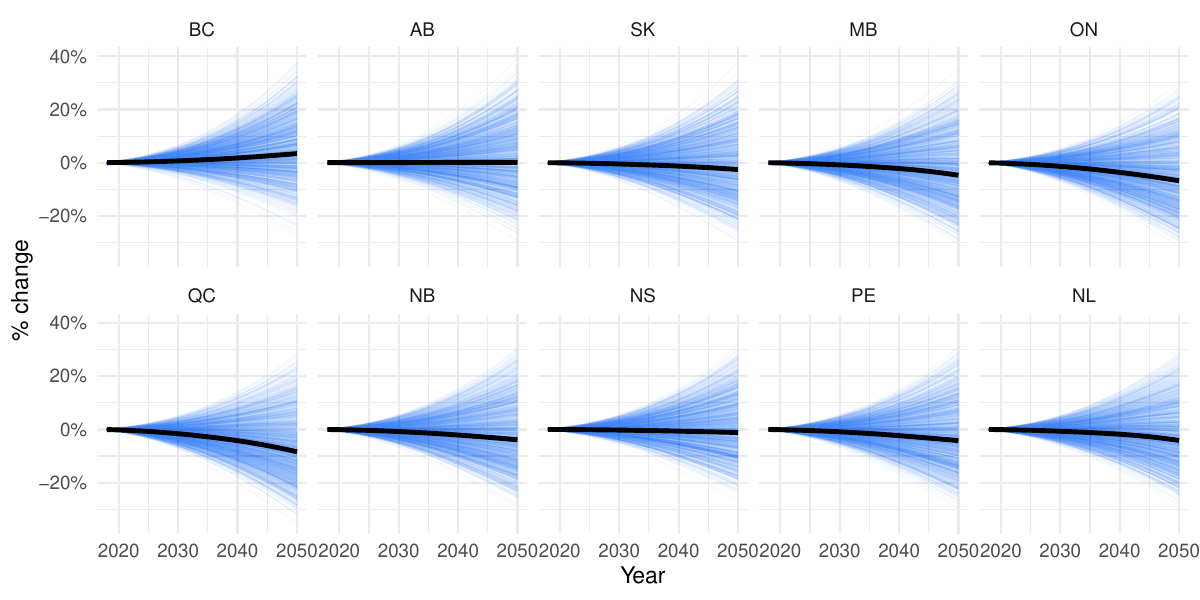}
         \caption{(k) Accommodation and Food Services}
    \end{subfigure} 
    \begin{subfigure}[b]{\linewidth}
         \centering
         \includegraphics[width=.9\linewidth]{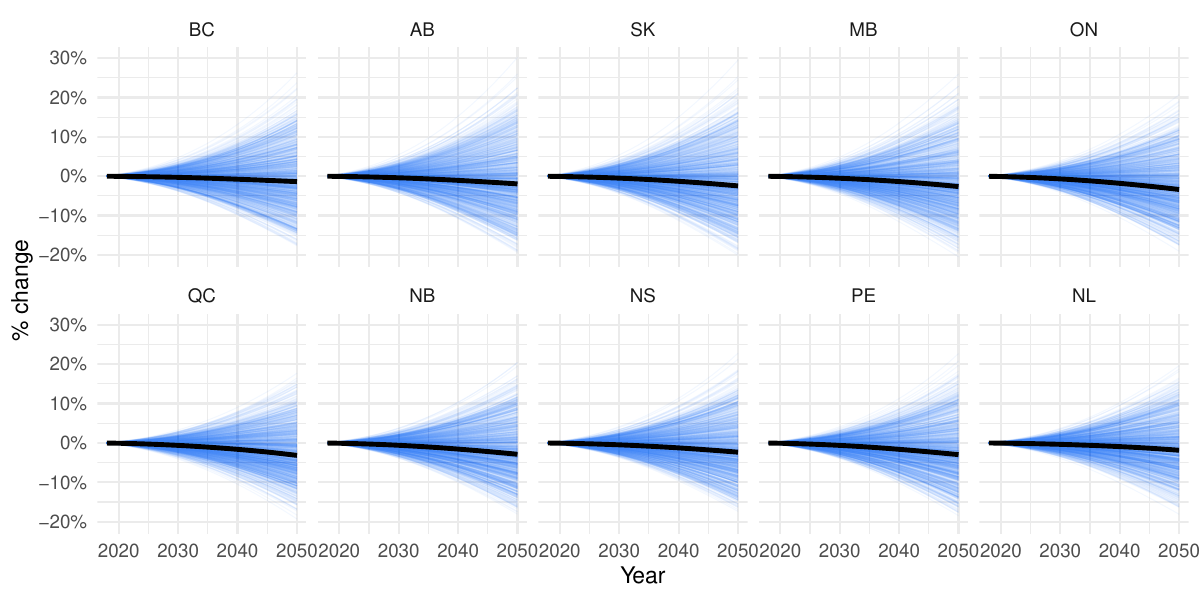}
         \caption{(l) Others}
    \end{subfigure} 
    \caption{Projected impact of climate change on GDP per capita across provinces and industries under RCP4.5 with uncertainty calculated using province-based block bootstrap.} 
\end{figure}

\begin{figure}[htbp]\ContinuedFloat
\captionsetup[subfigure]{labelformat=empty}
    \begin{subfigure}[b]{\linewidth}
         \centering
             \includegraphics[width=.9\linewidth]{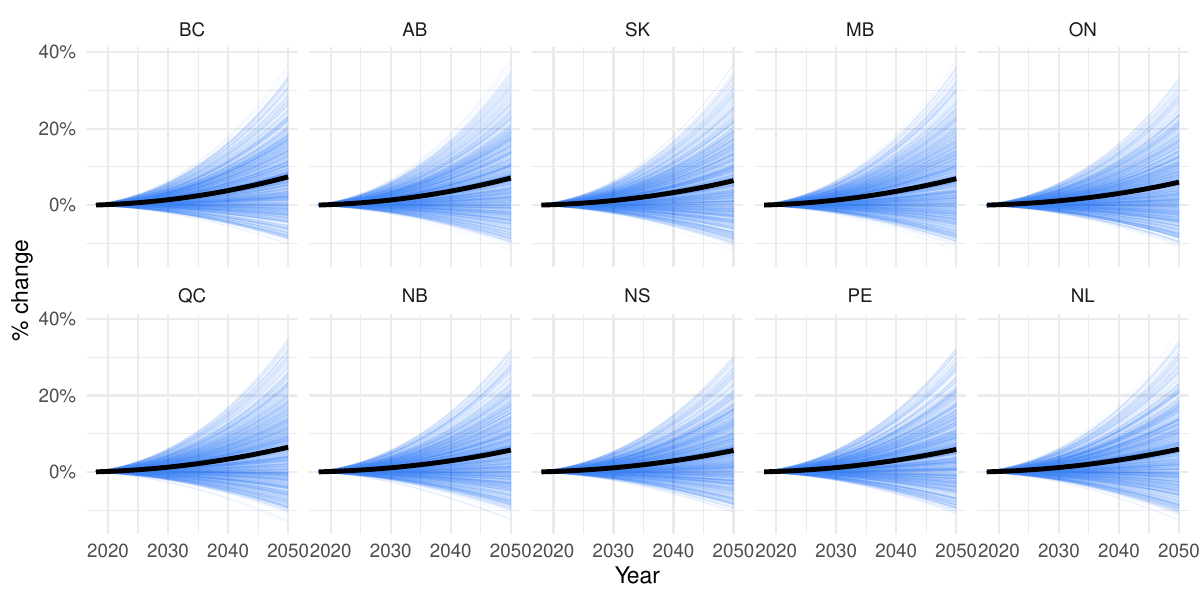}
         \caption{(m) Public Administration}
    \end{subfigure} 
    \begin{subfigure}[b]{\linewidth}
         \centering
         \includegraphics[width=.9\linewidth]{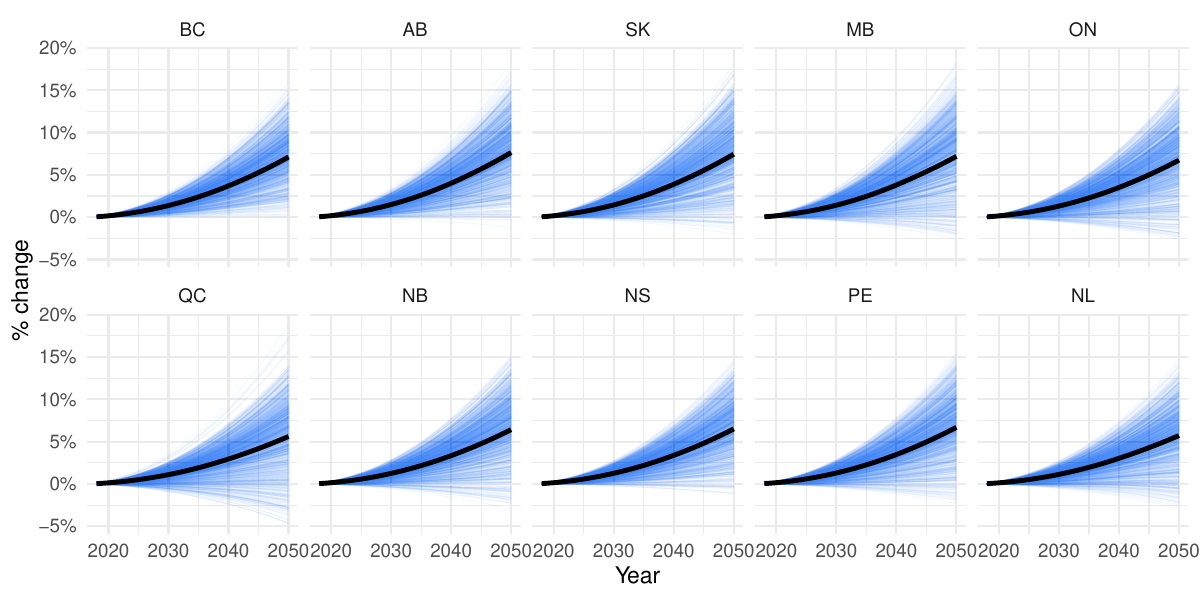}
         \caption{(n) Finance and Real Estate}
    \end{subfigure} 
    \begin{subfigure}[b]{\linewidth}
         \centering
         \includegraphics[width=.9\linewidth]{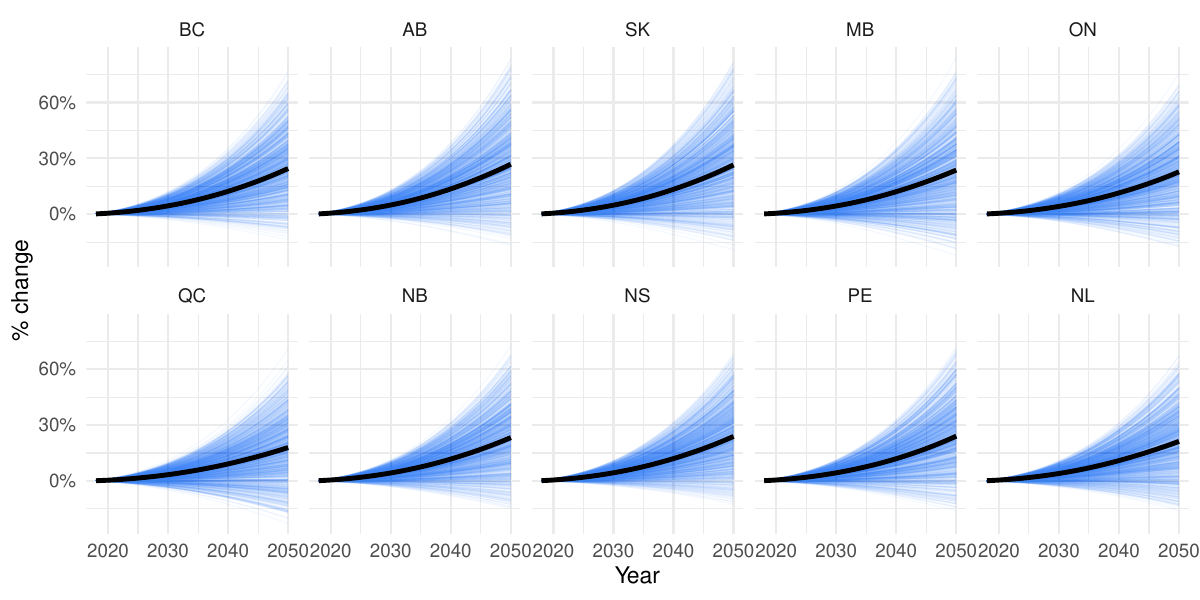}
         \caption{(o) Science and Technology}
    \end{subfigure} 
    \caption{Projected impact of climate change on GDP per capita across provinces and industries under RCP4.5 with uncertainty calculated using province-based block bootstrap.} 
\end{figure}

\FloatBarrier
\section{Sensitivity to Model Specification of Projection Results} \label{app:model-spec-proj}

\begin{figure}[H]
\captionsetup[subfigure]{labelformat=empty}
    \begin{subfigure}[b]{0.49\textwidth}
         \centering
         \includegraphics[width=.8\textwidth]{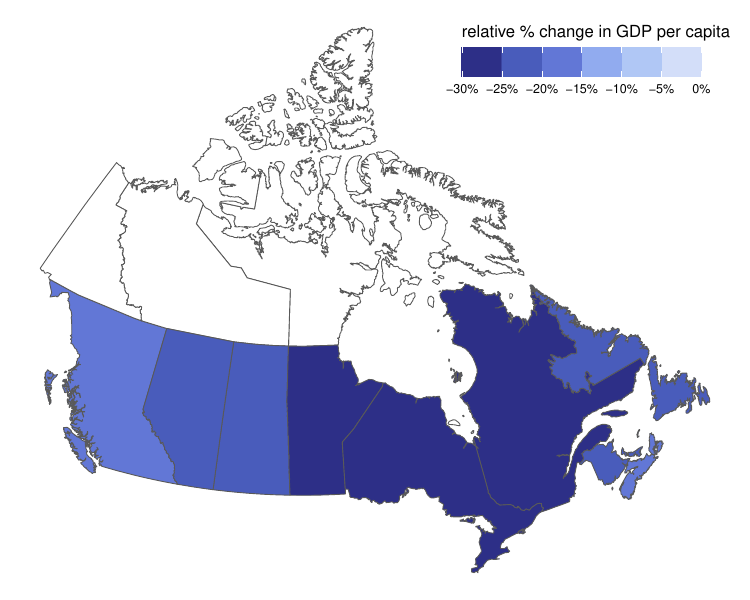}
         \caption{(1)}
    \end{subfigure} 
    \begin{subfigure}[b]{0.49\textwidth}
         \centering
         \includegraphics[width=.8\textwidth]{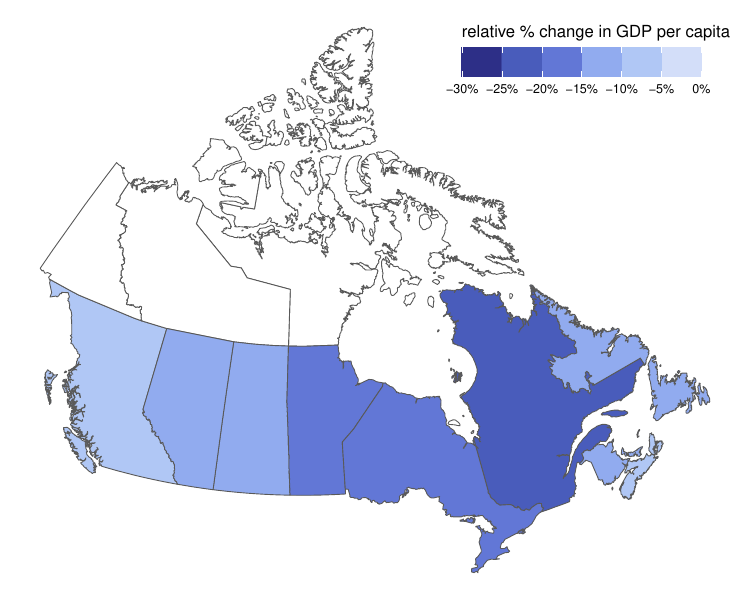}
         \caption{(2)}
    \end{subfigure}  
    
    \vspace{1cm}
    \begin{subfigure}[b]{0.49\textwidth}
         \centering
         \includegraphics[width=.8\textwidth]{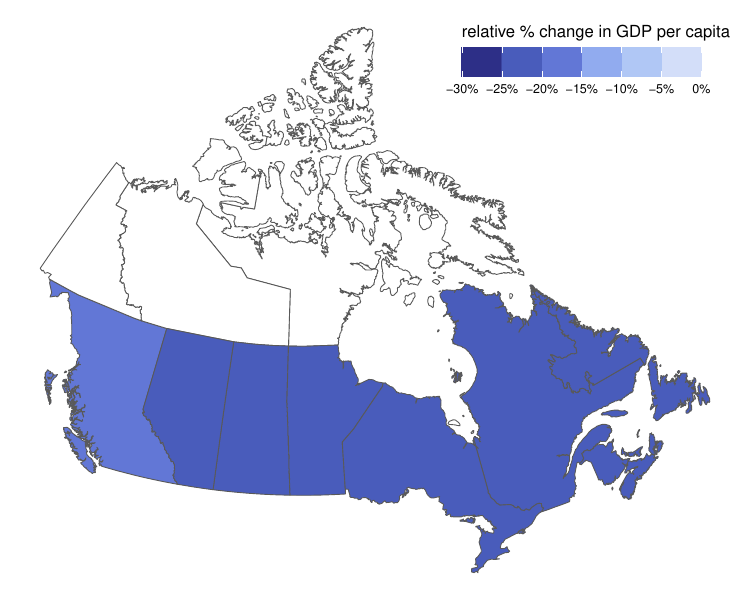}
         \caption{(3)}
    \end{subfigure}  
    \begin{subfigure}[b]{0.49\textwidth}
         \centering
         \includegraphics[width=.8\textwidth]{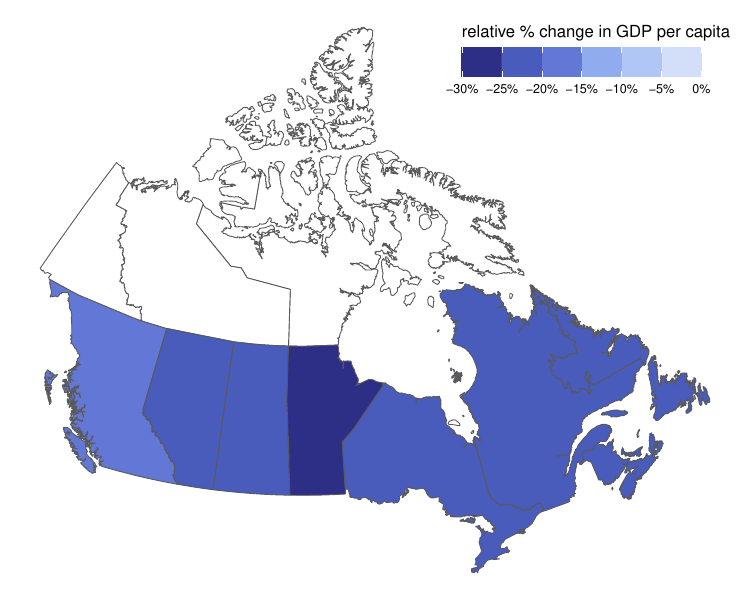}
         \caption{(4)}
    \end{subfigure} 
    
    \vspace{1cm}
    \begin{subfigure}[b]{0.49\textwidth}
         \centering
         \includegraphics[width=.8\textwidth]{report_fig/proj_canada_5.pdf}
         \caption{(5)}
    \end{subfigure}  
    \begin{subfigure}[b]{0.49\textwidth}
         \centering
         \includegraphics[width=.8\textwidth]{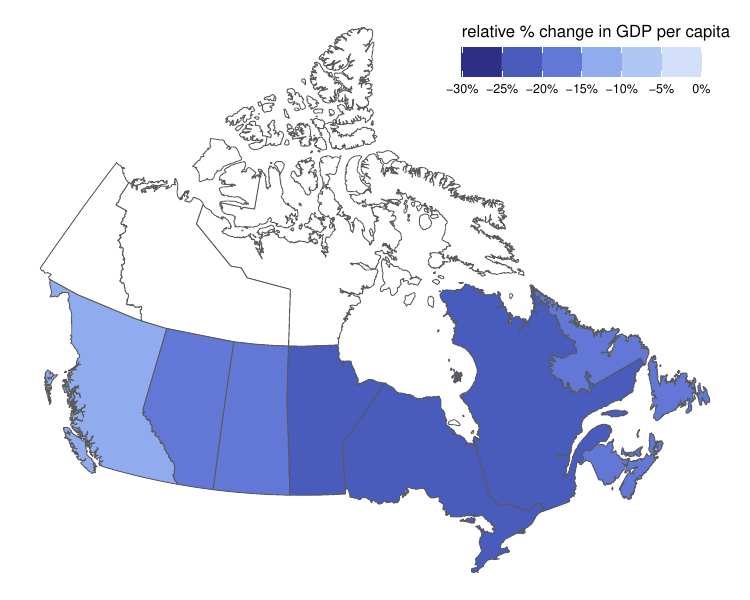}
         \caption{(6)}
    \end{subfigure}    
    \caption{Projected impacts of climate change on provincial GDP per capita by 2050 under RCP4.5 across different model misspecifications. }
    \label{fig:proj_gdp_other}
\end{figure}

\begin{figure}[H]
\captionsetup[subfigure]{labelformat=empty}
    \begin{subfigure}[b]{0.49\textwidth}
         \centering
         \includegraphics[width=.9\textwidth]{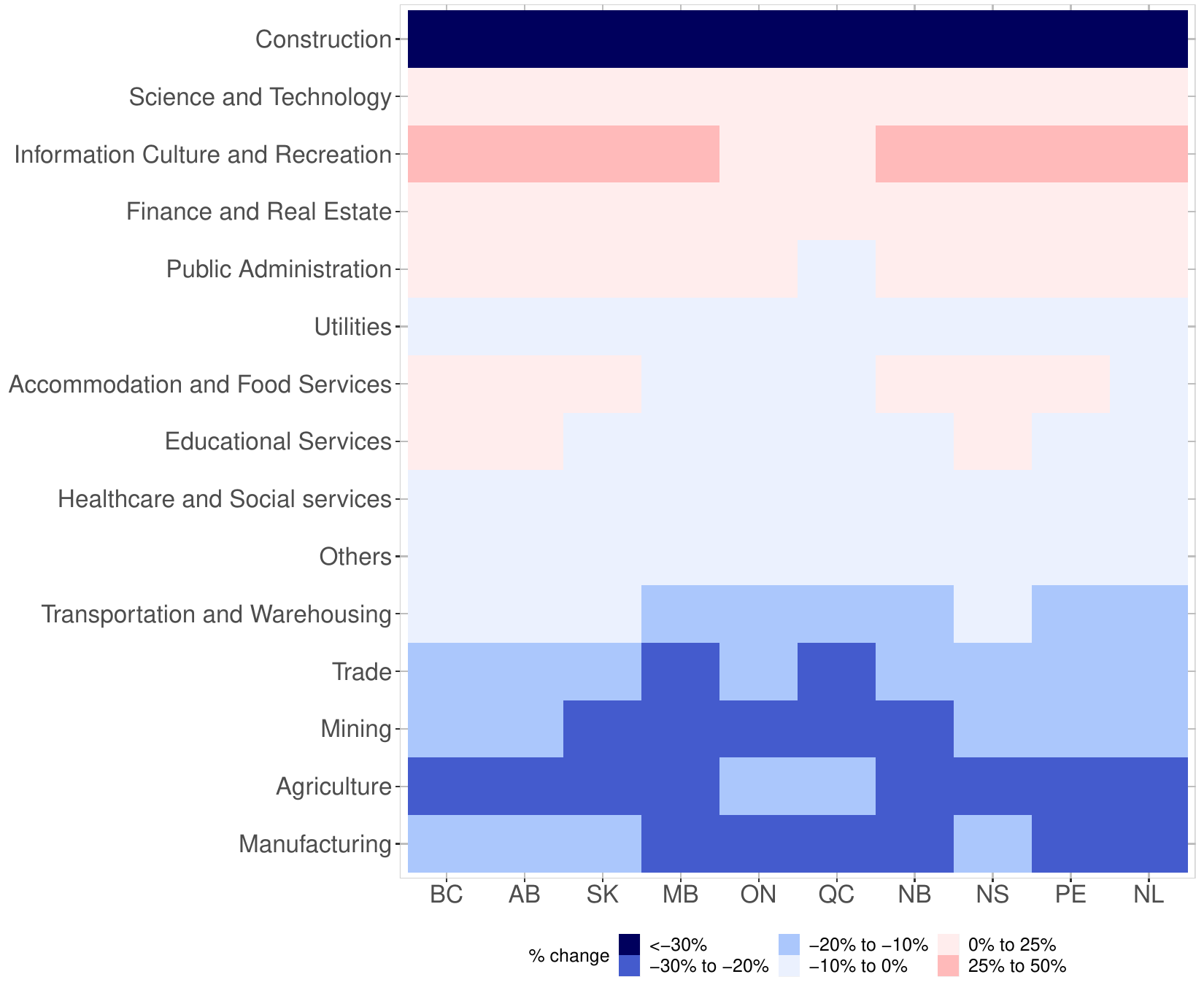}
         \caption{(1)}
    \end{subfigure} 
    \begin{subfigure}[b]{0.49\textwidth}
         \centering
         \includegraphics[width=.9\textwidth]{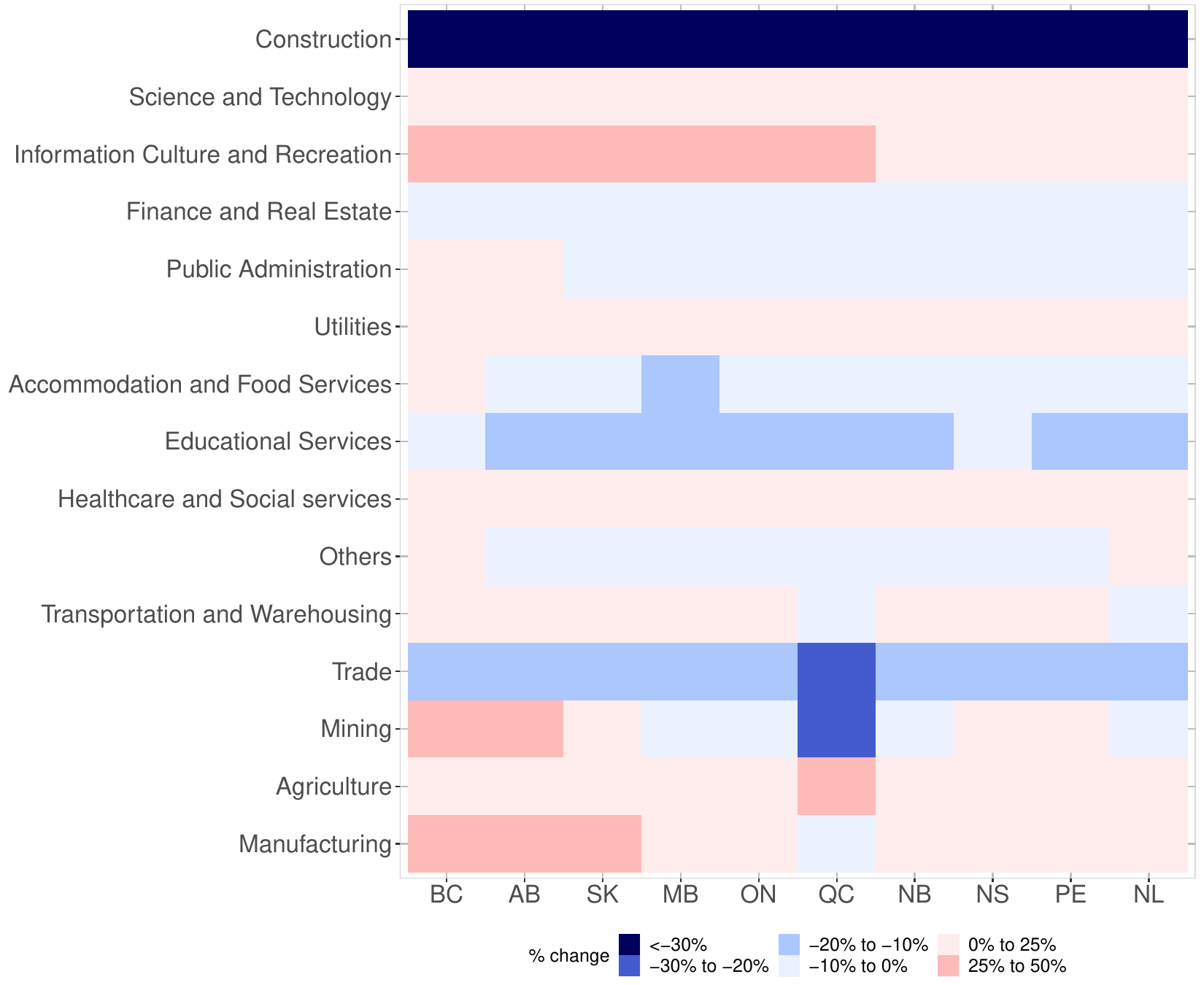}
         \caption{(2)}
    \end{subfigure}  
    
    \vspace{1cm}
    \begin{subfigure}[b]{0.49\textwidth}
         \centering
         \includegraphics[width=.9\textwidth]{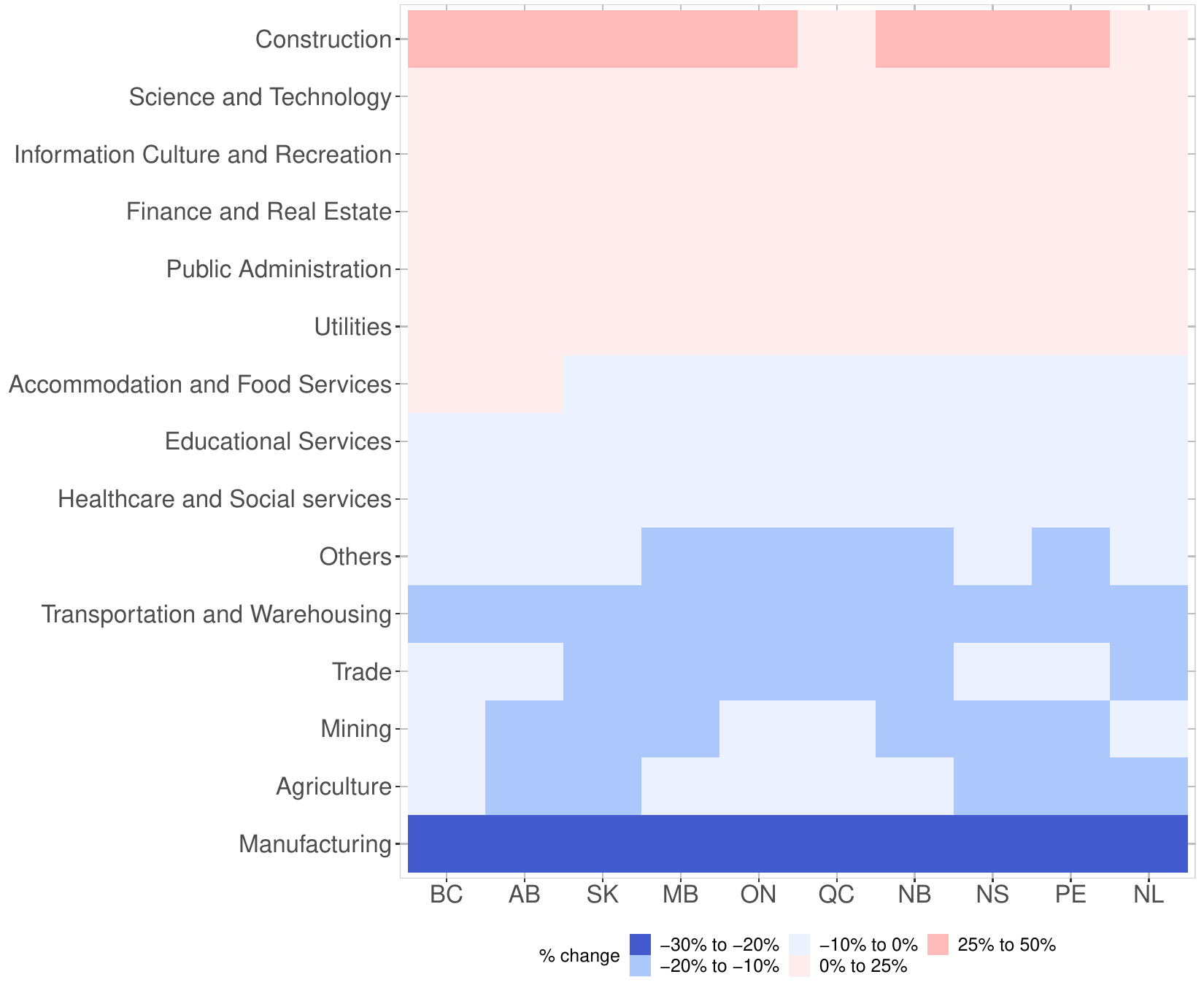}
         \caption{(3)}
    \end{subfigure}  
    \begin{subfigure}[b]{0.49\textwidth}
         \centering
         \includegraphics[width=.9\textwidth]{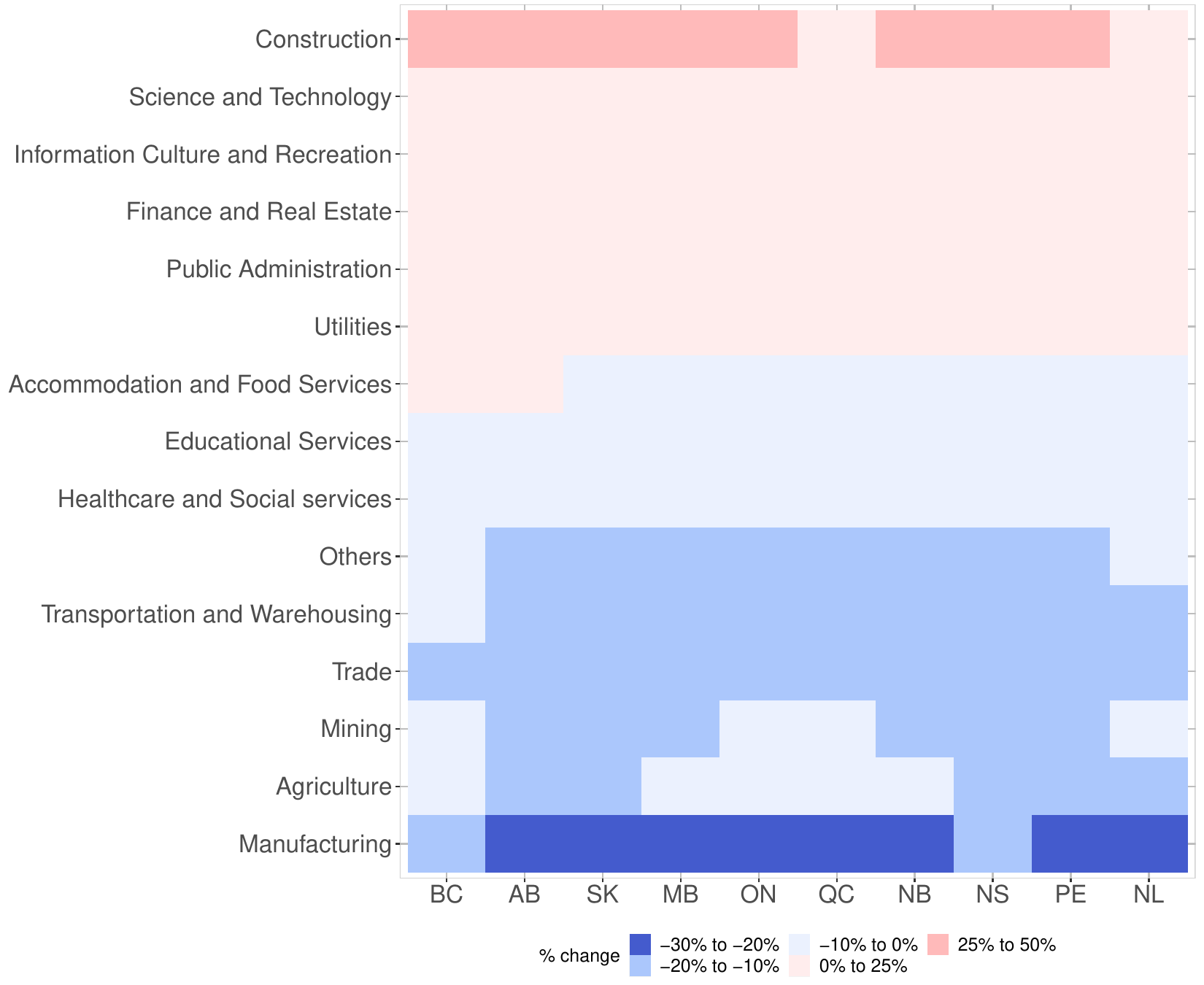}
         \caption{(4)}
    \end{subfigure} 
    
    \vspace{1cm}
    \begin{subfigure}[b]{0.49\textwidth}
         \centering
         \includegraphics[width=.9\textwidth]{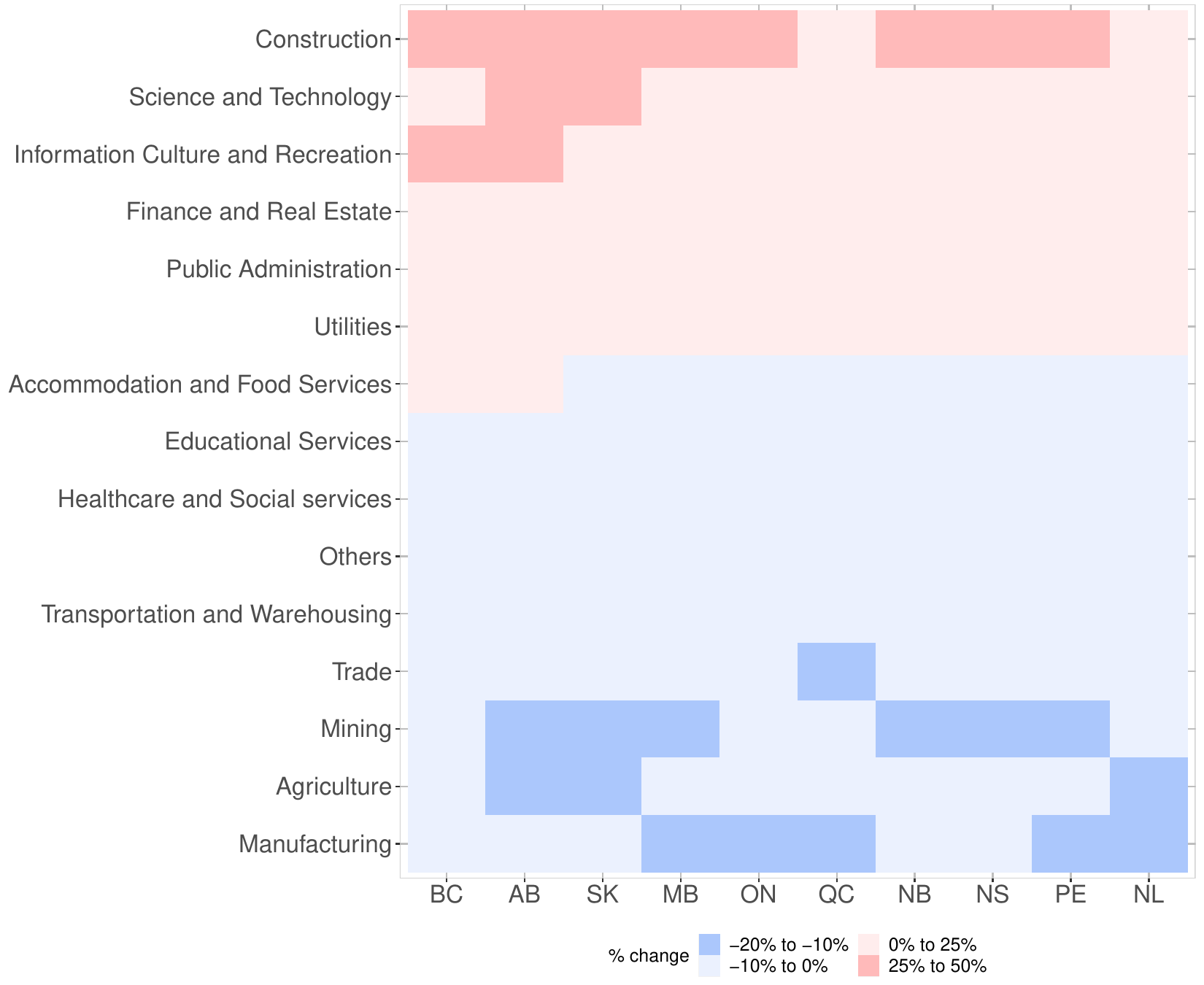}
         \caption{(5)}
    \end{subfigure}  
    \begin{subfigure}[b]{0.49\textwidth}
         \centering
         \includegraphics[width=.9\textwidth]{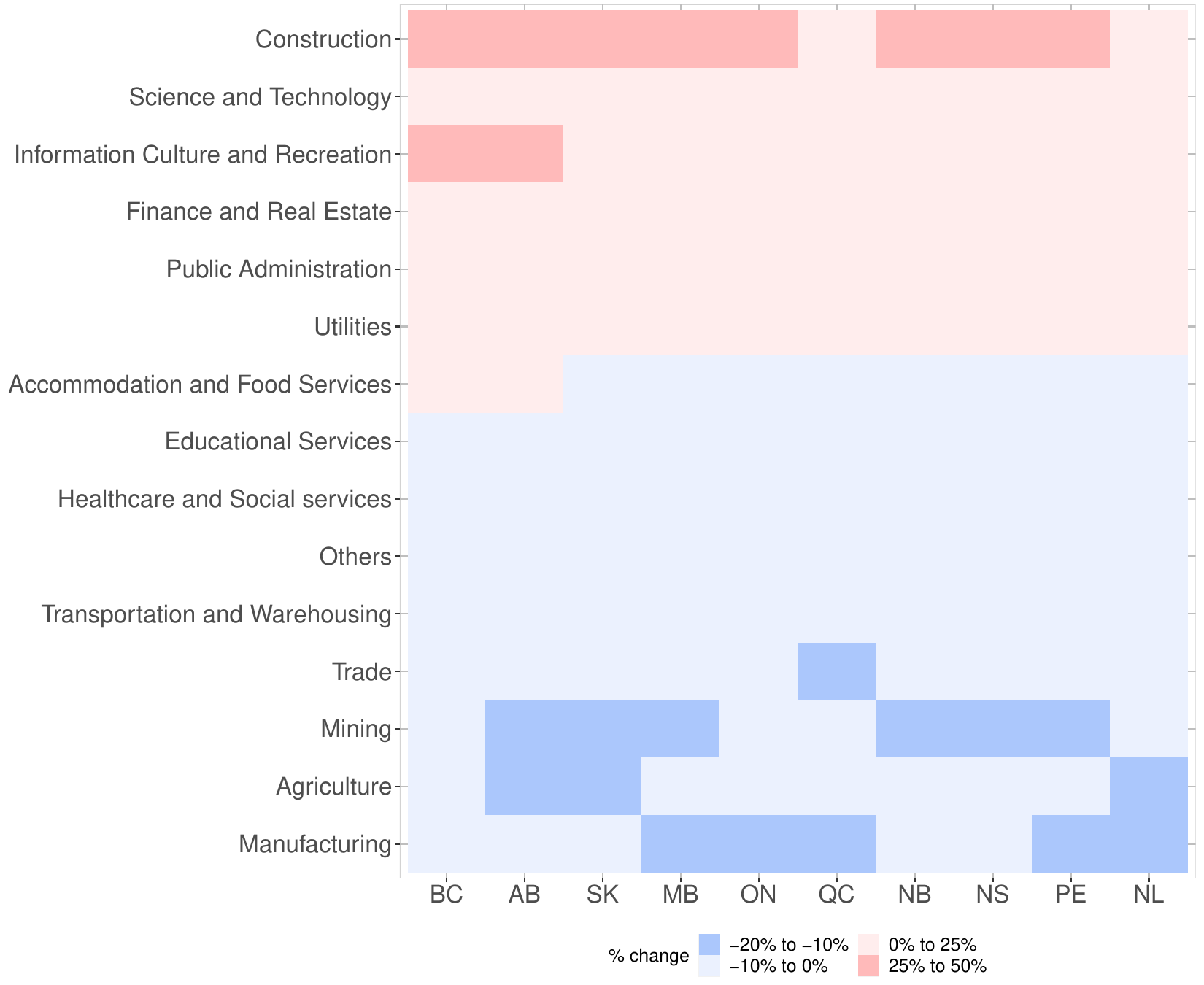}
         \caption{(6)}
    \end{subfigure}  
    \caption{Projected impacts of climate change on industry-specific GDP per capita by 2050 under RCP4.5 across different model misspecifications.}
    \label{fig:proj_ind_other}
\end{figure}

\section{Sensitivity Analysis of Climate Data}
\subsection{Weather Stations \label{app:weather-station}}

We obtained raw weather data from a total of 2,008 individual weather stations across Canada for the period of 1998 to 2017, with the number of weather stations per province ranging from 32 to 614. During the data cleaning process, we removed duplicated station IDs, and discarded records with missing or invalid coordinates. To ensure coverage, we computed the maximum number of months with complete records within each province for temperature and for precipitation, and retained stations with $\geq$ 90\% of that provincial maximum for both variables over the study period. As shown in Table~\ref{station_num}, this screening yields 412 qualified weather stations, with the provincial counts varying from 4 in Prince Edward Island and 90 in Qu\'ebec.

\begin{table}[ht]
\centering
\begin{tabular}{lrr}
\toprule
\textbf{Province} & \textbf{Before} & \textbf{After} \\ \midrule
Alberta & 614 & 57 \\
British Columbia & 266 & 82 \\
Manitoba & 189 & 40 \\
New Brunswick & 60 & 7 \\
Newfoundland & 128 & 29 \\
Nova Scotia & 81 & 12 \\
Ontario & 224 & 36 \\
Prince Edward Island & 32 & 4 \\
Quebec & 200 & 90 \\
Saskatchewan & 214 & 55 \\
\bottomrule
\end{tabular}
\caption{Station counts by province before and after data cleaning for the study period 1998-2017. }
\label{station_num}
\end{table}

\begin{figure}
    \centering
    \includegraphics[width=1.2\linewidth]{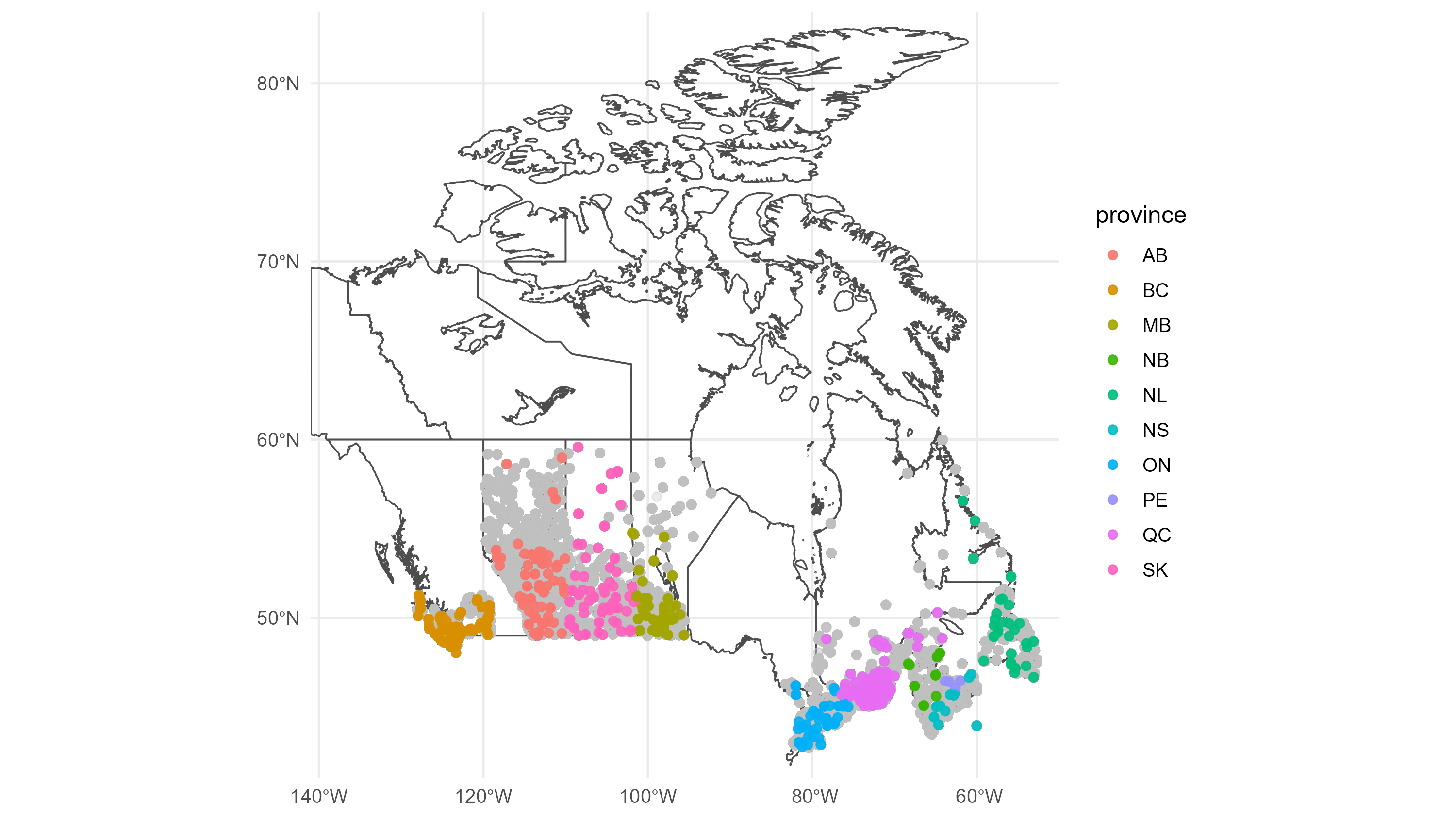}
    \caption{Spatial distribution of Canadian weather stations included in this study over the period 1998-2017. Each dot represents an individual station, with colors indicating the provinces. Grey dots denote stations that were excluded during the data cleaning process.}
    \label{fig:station_dist}
\end{figure}

Figure \ref{fig:station_dist} shows the spatial distribution of weather stations across Canada's ten provinces before and after data cleaning. Excluded stations are shown in grey, while retained stations are color-coded by province. Overall, the retained stations provide coverage comparable to that of the full station network in most provinces, with the exception of Alberta. Many northern stations in Alberta were excluded due to data completeness issues, which may introduce bias into the aggregated weather data (e.g., average temperatures may be biased upward because only southern stations are retained). However, we believe that the weather anomalies employed in our study, defined as deviations of seasonal weather measurements from long-term norms, are less affected by this potential bias.

Retained stations are highly clustered in the southern parts of the provinces, with coverage concentrated south of 55\degree N. Dense population of stations appears in southwestern British Columbia (Lower Mainland/Vancouver Island), across the southern Prairies (Alberta-Saskatchewan-Manitoba), and along the Ontario to Qu\'ebec corridor from Windsor through the Greater Toronto Area to southern Qu\'ebec. The Atlantic provinces exhibit strong coastal coverage (Nova Scotia, New Brunswick, Prince Edward Island, and the island of Newfoundland). Overall, the spatial coverage emphasizes settled and populated regions, with limited representation in northern and remote inland areas.

\subsection{Sensitivity Analysis Using Population-Weighted Climate Data} \label{app:sensitive-model-climate}

To assess the robustness of our results, we re-estimate all models in Table~\ref{tab:model_selection} using alternative constructions of the weather anomalies, and report the results in Table~\ref{app:anomaly-refit}. We calculate weather anomalies using population-weighted weather measurements in comparison to directly using the original weather measurements in the main paper. First, we clean the weather data following the same procedure. Next, to construct population-weighted weather anomalies, we match the GeoUID of each retained weather station to its corresponding sub-region based on the station’s geographic coordinates (i.e., longitude and latitude). Population data are then obtained from the 2016 Census. For each province, the weight assigned to a station is defined as the population of its sub-region divided by the total population represented by all retained stations within that province. These weights are used to compute weighted averages of temperature and precipitation at the provincial level. Finally, population-weighted anomalies are derived using Eq. (\ref{eq:temp_anomaly}) and Eq. (\ref{eq:prec_anomaly}), applied to the population-weighted temperature and precipitation data.

Using the population-weighted climate data to fit the six models presented in the main paper, we observe consistent patterns in Table~\ref{app:anomaly-refit}: negative impacts of Winter temperatures remain significant in four of the six models. Most precipitation coefficients are small and statistically insignificant across the two tables, which aligns with the findings of \citet{kahn2021}. Overall, this sensitivity analysis thus highlights the robustness of our results to the alternative calculation of climate variables.

\begin{table}[h]
\scriptsize
\renewcommand{\arraystretch}{1.1}
\centering
\begin{tabular}{lcccccc}
  \hline
  & (1C) & (2C) & (3C) & (4C) & (5C) & (6C) \\ 
  \hline
  \textbf{Model specification} \\
  Year Effect & Fixed & Fixed & No & No & Random & Random \\
  Economic Indices & No & No & Yes & Yes & Yes & Yes \\
  Major Economic Events & No & No & No & Yes & No & Yes \\
  \hline
  \textbf{Temperature} \\
  Spring Temp. & 0.0021  & 0.0008  & 0  & 0  & 0.0013  & 0.0013  \\ 
    & (0.0016) & (0.0024) & (0.0007) & (0.0007) & (0.0012) & (0.0012) \\ 
  Summer Temp. & 0.0011  & 0.003. & -0.0003  & -0.0003  & 0.0008  & 0.0008  \\ 
    & (0.0023) & (0.0016) & (0.0019) & (0.0019) & (0.0016) & (0.0016) \\ 
  Fall Temp. & -0.0034. & -0.004. & -0.0047. & -0.0047. & -0.0034  & -0.0034  \\ 
    & (0.0018) & (0.002) & (0.0022) & (0.0022) & (0.0021) & (0.0021) \\ 
  Winter Temp. & -0.0056* & -0.0053* & -0.0016  & -0.0016  & -0.0036* & -0.0036* \\ 
    & (0.0018) & (0.0018) & (0.0009) & (0.0009) & (0.0014) & (0.0014) \\ 
  (Spring Temp.)\verb|^|2 &  & -0.0009. &  &  &  &  \\ 
    &  & (0.0005) &  &  &  &  \\ 
  (Summer Temp.)\verb|^|2 &  & 0.0019. &  &  &  &  \\ 
    &  & (0.001) &  &  &  &  \\ 
  (Fall Temp.)\verb|^|2 &  & -0.0009  &  &  &  &  \\ 
    &  & (7e-04) &  &  &  &  \\ 
  (Winter Temp.)\verb|^|2 &  & -0.0001  &  &  &  &  \\ 
    &  & (0.0011) &  &  &  &  \\ 
  \hline
  \textbf{Precipitation} \\
  Spring Precip. & 0.0011  & 0.0052  & 0.0012  & 0.0012  & 0.0015  & 0.0015  \\ 
    & (0.0071) & (0.0075) & (0.0095) & (0.0095) & (0.0078) & (0.0078) \\ 
  Summer Precip. & 0.0065  & 0.0072  & 0.0024  & 0.0024  & 0.0045  & 0.0045  \\ 
    & (0.008) & (0.0086) & (0.0079) & (0.0079) & (0.0072) & (0.0072) \\ 
  Fall Precip. & 0.0064  & 0.008  & 0.0113  & 0.0113  & 0.0092  & 0.0092  \\ 
    & (0.0081) & (0.007) & (0.0085) & (0.0085) & (0.0081) & (0.0081) \\ 
  Winter Precip. & 0.0058  & 0.0042  & -0.0001  &  -0.0001  & 0.0024  & 0.0024  \\ 
    & (0.0031) & (0.0039) & (0.0051) & (0.0051) & (0.0042) & (0.0042) \\ 
  (Spring Precip.)\verb|^|2 &  & -0.0105. &  &  &  &  \\ 
    &  & (0.0052) &  &  &  &  \\ 
  (Summer Precip.)\verb|^|2 &  & 0.006  &  &  &  &  \\ 
    &  & (0.0068) &  &  &  &  \\ 
  (Fall Precip.)\verb|^|2 &  & -0.0067  &  &  &  &  \\ 
    &  & (0.0119) &  &  &  &  \\ 
  (Winter Precip.)\verb|^|2 &  & -0.0063  &  &  &  &  \\ 
    &  & (0.0131) &  &  &  &  \\ 
  \hline
  AIC & -832.9949 & -824.5437 & -818.3226 & -816.3226 & -825.2469 & -823.2469 \\ 
  BIC & -709.6079 & -675.1806 & -737.1470 & -731.8999 & -740.8243 & -735.5773 \\ 
   \hline
\end{tabular} 
\caption{Model sensitivity to climate calculation. The coefficients are rounded to the nearest 4 digits. Significant codes $***, **, *$, and $.$ correspond to significance levels 0.001, 0.01, 0.05, and 0.1, respectively. The unit of measurement for precipitation shown in this table is percentage points. The robust clustered standard deviations are given in brackets.}\label{app:anomaly-refit}
\end{table}

\end{document}